\documentclass[useAMS,usenatbib,usegraphicx]{mn2e}
\usepackage{ gensymb }

\usepackage{graphicx}
\usepackage[fleqn]{amsmath}
\usepackage{amssymb}

\newcommand{\p}{\partial}
\newcommand{\dd}{{\rm d}}
\newcommand{\M}{{\cal M}}
\newcommand{\F}{{\cal F}}
\newcommand{\sh}{{\rm sh}}
\newcommand{\rsh}{{r_{\rm sh}}}
\newcommand{\vort}{{\rm vort}}

\title[Angular momentum redistribution by SASI spiral modes]
{Angular momentum redistribution by SASI spiral modes and consequences for neutron star spins}

\author[Guilet \& Fern\'andez]
{J\'er\^ome Guilet$^{1,2}$ and Rodrigo Fern\'andez$^{3,4,5}$\\
$^1$ Department of Applied Mathematics and Theoretical Physics, University of Cambridge \\
 Centre for Mathematical Sciences, Wilberforce Road, Cambridge CB3 0WA, UK \\
$^2$ Max-Planck-Institut fur Astrophysik, Karl-Schwarzschild-Str. 1, D-85748 Garching, Germany \\
$^3$ Institute for Advanced Study, Princeton, NJ 08540, USA\\
$^4$ Department of Physics, University of California, Berkeley, CA 94720, USA\\
$^5$ Department of Astronomy \& Theoretical Astrophysics Center, University of California, Berkeley, CA 94720, USA
}

\begin{document}

\maketitle

\label{firstpage}

\begin{abstract}
In the collapsing core of massive stars, the standing accretion shock
instability (SASI) can drive spiral modes that efficiently redistribute angular
momentum. This process can impart a spin to the forming neutron star even when
the progenitor star is non-rotating.  Here we develop the first analytical
description of the angular momentum redistribution driven by a spiral mode of
the SASI. Our analysis, valid in the limit of small mode amplitude, shows that
the angular momentum separation is driven by the Reynolds stress generated by
the spiral mode. The resulting solutions compare favorably with previous
three-dimensional hydrodynamic simulations of the SASI in the linear and weakly
non-linear phases. Reasonable agreement is also found when extrapolating the solutions
into the fully non-linear phase. A Reynolds-decomposition of the flow is performed in the
saturated state of these simulations, showing that outward angular momentum
transport by the Reynolds stress and the fluctuating component of the mass flux balance inward
transport by advection. We derive an approximate analytic expression for the maximum angular momentum deposited in
the neutron star as a function of the mass accretion rate, shock radius, shock
compression ratio, and amplitude of the spiral mode at the time of explosion.
Implications for the birth spin periods of neutron stars are discussed.
\end{abstract}


%%%%%%%%%%%%%%%%%%%%%%%%%%%%%%%%%%%%%%%%%%%%%%%%%%%%%%%%%%%%%%%%%%%%%%%%%%%%%%%%%%
\section{Introduction}
Neutron stars are formed during the gravitational collapse of massive stars.
This core collapse also powers a supernova explosion that ejects
the outer layers of the progenitor. 
The dynamics of the explosion has important consequences for the 
properties of the resulting neutron star, such as its final mass, space velocity,
spin, and magnetic field (see, e.g., \citealt{janka2012a} for a review).

The most promising channel to drive an explosion for 
the majority of stellar progenitors is the so-called 
\emph{neutrino mechanism}, in which neutrino energy deposition
revives the stalled bounce accretion shock \citep{bethe85}. 
Breaking of the initial spherical symmetry is however key to a successful explosion 
\citep{liebendoerfer01,rampp02,thompson03,sumiyoshi05}.
Asymmetries are caused primarily by two hydrodynamical instabilities that 
operate in the region between the neutrinosphere and the shock:
neutrino-driven convection (e.g., \citealt{herant92}), and the Standing
Accretion Shock Instability (SASI; \citealt{blondin03}). Turbulence driven by these
instabilities increases the efficiency of neutrino energy deposition by increasing the
residency time of matter in the \emph{gain} region, where heating dominates
cooling, and by enlarging the postshock volume (e.g., \citealt{murphy08,marek09}).

The SASI is a global oscillatory instability driven by an unstable cycle of 
advective and acoustic perturbations that propagate between the shock and 
a region of strong deceleration close to the proto-neutron star surface \citep{foglizzo07,foglizzo09,guilet12}. 
For the conditions prevailing in core collapse supernovae,
the instability is dominated by low-frequency, large-scale modes with spherical 
harmonic indices $l \sim 1-2$ \citep{foglizzo07,yamasaki07}. 
In axisymmetric simulations, these modes manifest 
as sloshings of the shock along the axis. Relaxing the constraint of axisymmetry allows the existence
of spiral modes \citep{blondin07a}. 

In the absence of initial rotation in the collapsing core, spiral SASI modes have the
same linear growth rate as a sloshing mode with the same frequency and
spherical harmonic index $l$ \citep{foglizzo07}. 
A rotating progenitor favors the growth of prograde modes 
\citep{blondin07a,yamasaki08,iwakami09}. Hydrodynamic simulations without
neutrino heating have shown that even in the absence of rotation, the non-linear
evolution of SASI can lead to the dominance of a single spiral mode
\citep{blondin07a,fernandez10}. This numerical result has been confirmed
in an experimental analog of the SASI by \citet{foglizzo12}. 
The timescale over which the spiral mode becomes dominant,
and the conditions under which this happens are however not well understood. 
Furthermore, the interplay between the SASI and convection remains an
open area of research (e.g., \citealt{scheck08,iwakami08,fernandez09b,burrows12,mueller12,murphy2012,hanke13,iwakami13,fernandez2013}).

\cite{blondin07a} showed that SASI spiral modes have the ability to
redistribute angular momentum in the postshock region, and suggested that
this process could impart enough angular momentum to the neutron star to 
significantly change its spin. \cite{blondin07b} and \cite{fernandez10} have more closely examined the process
of angular momentum redistribution in spiral SASI flows, confirming the potential
for neutron star spin-up. \citet{foglizzo12} later confirmed a redistribution of angular momentum by spiral modes in their experimental analog of SASI. All existing studies of this process are based on numerical simulations or experiments, however,
and have not provided an analytical or physical description of this angular momentum 
separation. 

It is the purpose of this paper to provide such an analytical description. In pursuing
this, we set aside the broader question of the incidence of spiral modes in a realistic
supernova context, particularly in relation to neutrino-driven convection. Such a
topic is a current subject of debate in the literature (e.g., \citealt{janka2012b,burrows2013}), 
and progress on it requires work focused on the explosion dynamics.
Here we confine ourselves to flows in which convection is weak relative to the SASI (e.g., as
quantified by the ratio of advection time to convective growth time; \citealt{foglizzo06}; and
as seen in the models of \citealt{mueller12}),
and which are such that the SASI grows to reach a quasi-steady state (e.g., prior to
the onset of explosion). 
In addition to analytical studies, we carry out additional post-processing
of the three-dimensional hydrodynamic simulations of \citet{fernandez10}.

The paper is organised as follows. In Section~\ref{sec:formalism}, we derive a
formalism which allows to compute the radial profile of angular momentum
induced by a small amplitude spiral wave. These semi-analytical predictions are
then successfully compared to the 3D simulations of \citet{fernandez10} in
Section~\ref{sec:linear}. The radial profiles of angular momentum driven by
higher frequency harmonics display oscillations, which are explained in
Appendix~\ref{sec:decomposition} using a decomposition of the perturbations
into advected and acoustic waves. In Section~\ref{sec:nonlinear}, we analyse
the non-linear phase of the SASI. Finally, in
Section~\ref{sec:approximate} we derive an approximate analytical expression
for the angular momentum contained in a spiral wave and therefore for the
maximum angular momentum that can be imparted 
to the neutron star after the explosion has
succeeded. The results are discussed and conclusions are drawn in
Section~\ref{sec:conclusion}.

%%%%%%%%%%%%%%%%%%%%%%%%%%%%%%%%%%%%%%%%%%%%%%%%%%%%%%%%%%%%%%%%%%%%%%%%%%%%%%%%
\section{Formalism}
\label{sec:formalism}

We consider a standing spherical accretion shock around a central protoneutron
star of mass $M$ and radius $r_*$ that is subject to a  spiral SASI mode around some axis. The upstream accretion flow is non-rotating.
The system is described using spherical polar
coordinates $\{ r,\theta,\phi\}$ centred on the star. Following \cite{fernandez10}, we 
define the surface integrated angular momentum density along the symmetry axis ($z$) of the spiral
mode as
\begin{equation}
l_z(r,t) \equiv \iint r \sin\theta \rho v_\phi \, \dd^2 s = r^3\iint \rho v_\phi \sin\theta \, \dd \Omega,
	\label{eq:def_lz}
\end{equation}
where $\rho$ and $v_\phi$ are the fluid density and azimuthal velocity, respectively. This
expression is related to the total angular momentum $L_z$ enclosed in a spherical shell 
between radii $r_1$ and $r_2$ by
\begin{equation}
L_z(r,t) = \int_{r_1}^{r_2} l_z\, \dd r.
\end{equation}
Angular momentum conservation can then be written as
\begin{equation}
\p_t l_z + \p_r \F = 0,
	\label{eq:ang_mom_conservation}
\end{equation}
where $\F$ is the angular momentum flux integrated over a spherical surface
\begin{equation}
\F(r,t) \equiv  r^3\iint \rho v_r v_\phi \sin\theta \, \dd \Omega,
	\label{eq:def_angmom_flux}
\end{equation}
with $v_r$ the radial velocity.

We assume that the flow can be described as a stationary background with
superimposed small amplitude perturbations:
\begin{eqnarray}
\rho(r,\theta, \phi, t) & = & \rho_0(r) + \delta\rho(r,\theta, \phi, t) + \delta^2\rho(r,\theta, \phi, t) + ... \\
v_r & = & v_0 + \delta v_r + \delta^2 v_r + ... \\
v_\phi & = & \delta v_\phi + \delta^2 v_\phi + ...
\end{eqnarray}
where $\delta$ and $\delta^2$ denote first- and second order Eulerian perturbations, respectively, 
with $\delta \gg \delta^2$. We retain perturbations up to second order because this is the first non-zero
contribution to the surface integrated angular momentum $l_z$. The evolution of
the first order perturbations can be computed with a linear analysis as in
\cite{foglizzo07}. 

In the following we assume that first-order perturbations can be decomposed into a superposition of modes with a spherical harmonic angular 
dependence with indices $\{l,m\}$, and the time-dependence of a plane
wave with complex frequency $\omega=\omega_r + i\omega_i$, with $\omega_r$ and $\omega_i$ 
the real and imaginary parts, respectively. Except for transverse velocities, the  space and time dependence of
an arbitrary first-order perturbation $\delta A$ is
\begin{equation}
\delta A(r, \theta, \phi, t) = \sum_{l,m} Re\left[\delta \tilde{A}_{l,m}(r) e^{-i\omega t}Y_{l}^{m}(\theta, \phi)\right]
\end{equation}
where $\delta\tilde{A}_{l,m}(r) $ is the complex amplitude, and $Y_{l}^{m}$ is a complex spherical harmonic. 
Transverse velocities have a different spatial dependence, and the azimuthal velocity satisfies
\begin{equation}
\delta v_\phi(r, \theta, \phi, t) = \sum_{l,m}Re\left[\delta  \tilde{v}_{\phi,l,m}(r) e^{-i\omega t}\frac{im}{\sin\theta} Y_{l}^{m}(\theta, \phi)\right].
\end{equation}
Note that the surface integral of all these first order perturbations vanishes (as long as $l\neq 0$) :
\begin{equation}
\iint \delta A(r, \theta, \phi, t)\,\dd^2 s = 0.
\end{equation}
The second order perturbations are so far unknown and their surface integral does not necessarily vanish. 

Using the above decomposition, the surface integrated angular momentum density and flux can be expressed as
\begin{eqnarray}
l_z &=& -\frac{\dot{M}r}{4\pi}\iint \left\lbrack \frac{\delta \rho}{\rho_0}\frac{\delta v_\phi}{v_0} + \frac{\delta^2v_\phi}{v_0} \right\rbrack \sin\theta \, \dd\Omega , \\
\label{eq:ang_mom_flux_lz}
\F &=& l_z v_0 + T_{Rey} 
\end{eqnarray}
where $\dot{M}\equiv -4\pi r^2\rho_0v_0$ is the stationary mass flux, and
$T_{Rey}$ is the surface-integrated Reynolds stress associated with the SASI
modes, defined by
\begin{eqnarray}
T_{Rey}(r,t) & \equiv & \iint \rho_0 \delta v_r\delta v_\phi r\sin\theta \, \dd^2 s. \\
& = & -\frac{\dot{M}rv_0}{4\pi}\iint \frac{\delta v_r\delta v_\phi}{v_0^2} \sin\theta \, \dd\Omega.
	\label{eq:def_Trey}
\end{eqnarray}
The stress can be computed using the linear eigenmodes
\begin{eqnarray}
T_{Rey}&=& -\frac{\dot{M}rv_0}{4\pi} \iint Re\left(\sum_{l,m} \frac{\delta \tilde{v}_{\phi,l,m}}{v_0} e^{-i\omega_l t}\frac{im}{\sin\theta} Y_{l}^{m}\right) \nonumber\\
&& \times Re\left(\sum_{l^{\prime},m^{\prime}} \frac{\delta \tilde{v}_{r,l^{\prime},m^{\prime}}}{v_0} e^{-i\omega_{l^{\prime}} t}Y_{l^{\prime}}^{m^{\prime}}\right) \sin\theta \, \dd\Omega \label{eq:Trey0}
\end{eqnarray}
Using the relation $Re(z_1)Re(z_2) = (z_1+z_1^*)(z_2+z_2^*)/4 = \left[Re(z_1z_2^*) + Re(z_1z_2)\right]/2$, we obtain
\begin{eqnarray}
T_{Rey}&=& -\frac{\dot{M}rv_0}{8\pi}\sum_{l,m}\sum_{l^{\prime},m^{\prime}}Re\Big\lbrack \nonumber  \\
&& im \frac{\delta\tilde{v}_{\phi,l,m}}{v_0}\frac{\delta \tilde{v}_{r,l^{\prime},m^{\prime}}}{v_0} e^{-i(\omega_{l}+\omega_{l^{\prime}}) t}  \iint Y_{l}^{m}Y_{l^{\prime}}^{m^{\prime}} \, \dd\Omega \nonumber \\
&& +   im \frac{\delta\tilde{v}_{\phi,l,m}}{v_0}\frac{\delta \tilde{v}_{r,l^{\prime},m^{\prime}}^*}{v_0} e^{-i(\omega_{l}-\omega_{l^{\prime}}) t}  \iint Y_{l}^{m}Y_{l^{\prime}}^{m^{\prime}*} \, \dd\Omega  \Big\rbrack \label{eq:Trey1}\nonumber\\
\end{eqnarray}
The first term inside the brackets vanishes for the following reasons. Since $Y_l^{m*}=Y_l^{-m}$, we have $ \iint
Y_{l}^{m}Y_{l^{\prime}}^{m^{\prime}} \, \dd\Omega=0$ unless $\{l^\prime,m^\prime\}=\{l,-m\}$. 
This condition is met in two cases: $m=m^{\prime}=0$ for which the term
vanishes, or $m=-m^{\prime} $ for which the terms ($m,-m$) and ($-m,m$) cancel
each other. Then the second term of equation~(\ref{eq:Trey1}) can be simplified
using the relation $\iint Y_{l}^{m}Y_{l^{\prime}}^{m^{\prime}*} \,
\dd\Omega = \delta_{l,l^\prime}\delta_{m,m^\prime}$, with $\delta$
the Kronecker symbol.  We then obtain
\begin{equation}
T_{Rey}= -\frac{\dot{M}rv_0}{8\pi}\sum_{l,m}Re\Big\lbrack im \frac{\delta\tilde{v}_{\phi,l,m}}{v_0}\frac{\delta \tilde{v}_{r,l,m}^*}{v_0} e^{2\omega_{i,l} t}  \Big\rbrack. \label{eq:Trey2}
\end{equation}
Defining $T_{Rey0,l,m}$ as the Reynolds stress amplitude of a given mode 
with spherical harmonic indices $\{l,m\}$ and with the time dependence scaled out,
\begin{equation}
T_{Rey0,l,m}(r)= -\frac{\dot{M}rv_0}{8\pi} Re\Big\lbrack im \frac{\delta\tilde{v}_{\phi,l,m}}{v_0}\frac{\delta \tilde{v}_{r,l,m}^*}{v_0} \Big\rbrack \label{eq:Trey0lm},
\end{equation}
we can write equation~(\ref{eq:Trey2}) as
\begin{equation}
T_{Rey}= \sum_{l,m} T_{Rey0,l,m}(r) e^{2\omega_{i,l} t} \label{eq:Trey3}.
\end{equation}

Combining equations (\ref{eq:ang_mom_conservation}), (\ref{eq:ang_mom_flux_lz}), 
and (\ref{eq:Trey3}), we can write angular momentum conservation
as a partial differential equation for the evolution of $l_z$
\begin{equation}
\p_t l_z + \p_r(l_z v_0) = - \sum_{l,m} \p_r T_{Rey0,l,m}e^{2\omega_{i,l} t}.
	\label{eq:lz_PDE}
\end{equation}
The angular momentum density $l_z$ can therefore be 
written as the sum of contributions from different spherical harmonics, with each term given by 
\begin{equation}
\p_t l_{z,l,m} + \p_r(l_{z,l,m} v_0) = - \p_r T_{Rey0,l,m}e^{2\omega_{i,l} t}.
	\label{eq:lzlm_PDE}
\end{equation}

To solve equation~(\ref{eq:lzlm_PDE}), we look for a solution having the 
same time dependence as the Reynolds stress,
\begin{equation}
l_{z,l,m}(r,t) = l_{z0,l,m}(r)e^{2\omega_{i,l} t}.
\end{equation}
The partial differential equation is then reduced to an ordinary one:
\begin{equation}
\frac{\dd}{\dd r}(l_{z0,l,m}v_0) + \frac{2\omega_{i,l}}{v_0}l_{z0,l,m}v_0 = -\frac{\dd}{\dd r}(T_{Rey0,l,m}).
\end{equation}
The spherically-integrated angular momentum density of a single mode
is then
\begin{equation}
l_{z0,l,m} = -\frac{T_{Rey0,l,m}}{v_0} + \frac{e^{-2\omega_{i,l} \tau_{\rm adv}}}{v_0}\int_{\rsh}^r\frac{2\omega_i e^{2\omega_{i,l} \tau_{\rm adv}}}{v_0}T_{Rey0,l,m}\, \dd r, 
	\label{eq:lzlm}
\end{equation}
where $\tau_{\rm adv}(r)=\int_{\rsh}^r \dd r/v_0$ is the advection time from the shock radius $\rsh$ to a 
radius $r < \rsh$, and where we have used the following boundary condition at the shock:
\begin{equation}
\F(\rsh) = l_{z0}(\rsh)v_\sh + T_{Rey0}(\rsh) = 0,
	\label{eq:flux_shock}
\end{equation}
which follows from the vanishing angular momentum flux above the shock.
The first term in Equation~(\ref{eq:lzlm}) describes a situation where the mode
would not grow, while the second term, proportional to the growth rate, is a
correction that takes into account the time dependence. 

Note that a sum over spherical harmonics indices should be performed in order
to obtain the total angular momentum density. Also note that while the angular
momentum driven by spirals with different spherical harmonics indices simply
adds up, the same is not true for several spiral modes with different
frequencies but the same spherical harmonics indices. In the latter case, cross
terms which are oscillatory in time would appear and would require a separate 
treatment which we do not provide here. A short discussion of the effect of
higher frequency harmonics on the angular momentum redistribution is 
given in Section~\ref{sec:r2}.

%-----------------------------------------------------------------------------
\subsection{Angular momentum density below the shock}

The angular momentum density below the shock due to a mode with
spherical harmonic indices $\{l,m\}$ follows from equations~(\ref{eq:Trey0lm}) and (\ref{eq:lzlm}),
\begin{equation}
l_{z\sh} = -\frac{T_{Rey0}(\rsh)}{v_\sh} = \frac{\dot{M}\rsh}{8\pi} Re\left(\frac{im\tilde{\delta v_\phi}\tilde{\delta v_r}^*}{v_0^2} \right)_\sh \label{eq:lzsh0}
\end{equation}
This expression can be evaluated using the boundary conditions for the linear eigenmodes. The complex amplitude of the azimuthal velocity perturbation
$\delta\tilde{v}_\phi$ is \citep[e.g.][]{guilet12}:
\begin{equation}
\left(\frac{\delta \tilde{v}_\phi}{v}\right)_\sh = \frac{v_1-v_\sh}{v_\sh} \frac{\Delta r}{r_\sh}, 
	\label{eq:dvphish}
\end{equation}
where $v_1$ and $v_\sh$ are the radial velocities upstream and
downstream of the shock, respectively. Given that this amplitude is real, only the imaginary
part of the amplitude of the radial velocity perturbation matters for the
angular momentum density:
\begin{equation}
Im\left(\frac{\delta \tilde{v}_r}{v}\right)_\sh = -\frac{\omega_r r_\sh}{v_\sh} \left(1 -1/\kappa \right)\frac{1+1/\M_1^2}{\gamma - (\gamma+1)/\kappa + 1/\M_1^2}\frac{\Delta r}{r_\sh},
	 \label{eq:dvrsh}
\end{equation}
where $\kappa\equiv v_1/v_\sh$ is the compression ratio of the shock, $\gamma$
is the adiabatic index of the gas, and $\M_1$ is the upstream Mach number. 
Note that this equation is valid for a shock with a constant dissociation energy 
(e.g., \citealt{fernandez09a}).

\begin{figure}
\centering
 \includegraphics[width=\columnwidth]{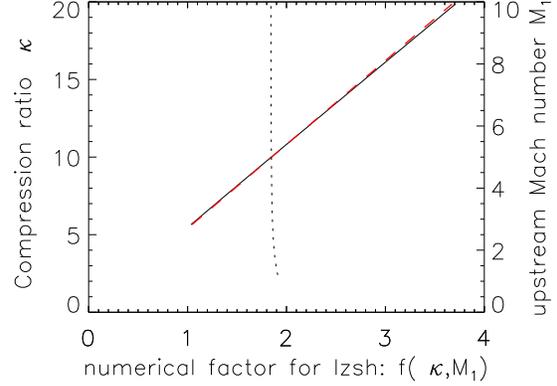}
 \caption{
Dependence of the numerical factor $f(\kappa,\M_1)$ that enters the
angular momentum density (equation~\ref{eq:lzsh}) on the shock compression
ratio $\kappa$ (left axis, full line) and on the upstream Mach number $\M_1$ (right axis, dotted line). When varying one parameter we fixed the other one to $M_1=5$ and $\kappa=10$, respectively (note that the shock dissociation energy must be varied for one
of the parameters to remain constant). The red dashed line shows the fit $f = 0.185\kappa$.}
\label{fig:lzsh_factor}
\end{figure}

Combining Equations~(\ref{eq:lzsh0}), (\ref{eq:dvphish}) and (\ref{eq:dvrsh}),
we can write
\begin{equation}
\frac{l_{z\sh}}{\dot{M}\rsh} = - m\frac{\omega_r\rsh}{2\pi v_\sh}f(\kappa,\M_1)\left(\frac{\Delta r}{\rsh}\right)^{2},
	\label{eq:lzsh}
\end{equation}
where $f(\kappa,\M_1)$ is a dimensionless factor,
\begin{equation}
f(\kappa,\M_1) \equiv \frac{1}{4}\left(\kappa -1\right)(1-1/\kappa)\frac{1+1/\M_1^2}{\gamma - (\gamma+1)/\kappa + 1/\M_1^2}.
\end{equation}
For a strong ($\M_1\rightarrow \infty$) adiabatic shock this expression yields $f=1/(\gamma^2-1)
=1.29$ (with $\gamma=4/3$), while for an adiabatic shock with $\M_1=5$ as
studied in Section~\ref{sec:linear}, we obtain $f =1.04$. 
The dependence of the numerical
factor $f$ on the compression ratio $\kappa$ and the upstream Mach number
$\M_1$ is shown in Figure~\ref{fig:lzsh_factor} (the dissociation energy is
varied when keeping one parameter constant). The dependence on $\M_1$ for a
fixed $\kappa$ is very weak; it is even exactly independent of $M_1$ if
$\kappa=(\gamma+1)/(\gamma-1)$, i.e., if the compression ratio is that of a
strong adiabatic shock. Interestingly, the angular momentum density below the
shock increases with $\kappa$, and the dependence is very close to linear, with
$f(\kappa,\M_1) \simeq 0.185\kappa$ providing a very good fit when
$\gamma=4/3$. A typical value of the compression ratio in the context of
supernovae $\kappa\sim 10$ yields $f=1.85$, which is somewhat larger than
for an adiabatic shock. Note also that the angular momentum density has 
the same sign as $m$ since $v_\sh$ is a negative quantity.

%%%%%%%%%%%%%%%%%%%%%%%%%%%%%%%%%%%%%%%%%%%%%%%%%%%%%%%%%%%%%%%%%%%%%%%%%%%%%%%%%
\section{Angular momentum profile in the linear phase}
	\label{sec:linear}

Here we use the formalism of Section~\ref{sec:formalism} to compute
the angular momentum profile during the linear phase of the SASI, and compare the
results with the 3D numerical simulations of \citet{fernandez10}. The linear
eigenmodes -- computed numerically as in \cite{guilet12} --
allow the construction of the Reynolds stress profile of each mode using
equation~(\ref{eq:Trey0lm}). The contribution of a given mode to the angular
momentum density profile is then obtained from equation~(\ref{eq:lzlm}). The amplitude of each of the spiral modes contributing to these profiles is extracted from the simulations as described in Appendix~\ref{sec:spiral_mode_amplitude}. Two methods have been tried, which rely on a fit of the time-evolution of either the shock displacement or the transverse velocities at a given radius below the shock. In the rest of this section, the mode amplitude is extracted from the transverse velocities at a radius $r=0.8r_{\rm s0}$ (Appendix~\ref{sec:fit_transverse_velocities}), as it was found to give more accurate results.

The simulations of \citet{fernandez10} evolve the hydrodynamic equations
with an initial condition equal to the stationary accretion shock system
described in \S\ref{sec:formalism}. 
The properties of the models studied are summarized in Table~\ref{tab:lz_tot}. 
Models are labeled by the ratio of the stellar to shock radius $r_*/r_{\rm s0}$,
which determines which modes become unstable when all other parameters are
kept constant\footnote{All models have upstream Mach number $\M_1=5$, adiabatic
index $\gamma=4/3$, vanishing energy flux upstream of the shock, and a cooling
function ${\cal L} \propto \rho P^{3/2}$. The normalization of the latter is
chosen so that the radial velocity vanishes at $r=r_*$.}. 
A given spiral mode is excited by placing an overdense shell with 
the same angular dependence in the
upstream flow. We divide the
rest of the discussion according to the type of mode in question: fundamental $l=1$ spiral 
mode (model R5\_L11\_HR), 
fundamental $l=2$ spiral mode (model R6\_L22\_P2),
and a model with multiple unstable $l=1$ harmonics (R2\_L11h).

Unless otherwise noted, the system of units is based on the initial shock
radius $r_{\rm s0}$, free-fall velocity at the shock $\sqrt{2GM/r_{\rm s0}}$, and upstream
density $\rho_1$. For an easier connection with physical supernova parameters, however, 
we normalize the spherically-integrated
Reynolds stress and angular momentum density by $\dot{M}r_{\sh0} |v_{\sh0}|$
and $\dot{M}r_{\sh0}$, respectively, where $v_{\sh0}$ is
the downstream shock velocity.

\begin{table*}
 \centering
 \begin{minipage}{130mm}
\caption{Summary of 3D models of \citet{fernandez10} with various derived quantities. Columns show model name,
initial ratio of stellar to shock radius, indices of spiral mode excited $\{l,m\}$, ratio of mode 
frequency to advection frequency below the shock,
amplitude of the mode in the saturated state, time interval for time average, and 
time-averaged angular momentum density from the simulation
as well as from the analytic estimate in Section~\ref{sec:approximate}}
  \begin{tabular}{@{}lrrrrrrr@{}}
  \hline
   Simulation     & $r_*/r_{\rm s0}$   &  $\{l,m\}$  & $\frac{\omega_r(\rsh-r_*)}{(2\pi |v_\sh|)}$ 
                  & Amplitude & time interval &  $l_z/(\dot{M}r_\sh^2)$  & $l_z/(\dot{M}r_\sh^2)$   \\
    name          &       &            &    
                  &           &               & simulation               &  analytic \\
 \hline
 R5\_L11\_HR      & 0.5 & 1,1  &  0.39 & 1.03  & [75,125] & 0.64   & 0.43   \\ 
 R6\_L22\_P2      & 0.6 & 2,2  &  0.44 & 0.36 &  [50,100] & 0.21   & 0.12   \\
 R2\_L11h         & 0.2 & 1,-1 &  0.42 & 1.31 &  [75,125] & -0.80  & -0.76  \\
\noalign{\smallskip}
 R6\_L21\_P2      & 0.6 & 2,1 &  0.44 & 0.41 &  [50,100] & 0.092 & 0.077 \\
 R2\_L11f         & 0.2 & 1,1 &  0.42 & 1.16 &[150,200] & 0.74   & 0.59  \\
\hline
\end{tabular}
\end{minipage}
\label{tab:lz_tot}
\end{table*}

%---------------------------------------------------------------
\subsection{Fundamental $l=1$ spiral mode}
	\label{sec:r5}

Choosing $r_*/r_{\rm s0}=0.5$ results in a configuration with an unstable fundamental $l=1$ mode with
no unstable harmonics, and hence model R5\_L11\_HR is initially perturbed to excite the $\{l,m\}=\{1,1\}$ spiral.
In order to compare directly the semi-analytical results with the simulations,
we first extract the amplitude of the two $m=\pm 1$ spiral
modes using the method described in Appendix~\ref{sec:fit_transverse_velocities}. The resulting amplitudes of the two spiral modes $m=1$ and $m=-1$ at
time $t=30$ are $A_1=0.229$ and $A_{-1}=0.055$, respectively. While the $m=1$ spiral is the dominant mode, the $20\%$ contribution from the $m=-1$ mode must be included to improve accuracy.

\begin{figure*}
\centering
 \includegraphics[width=\columnwidth]{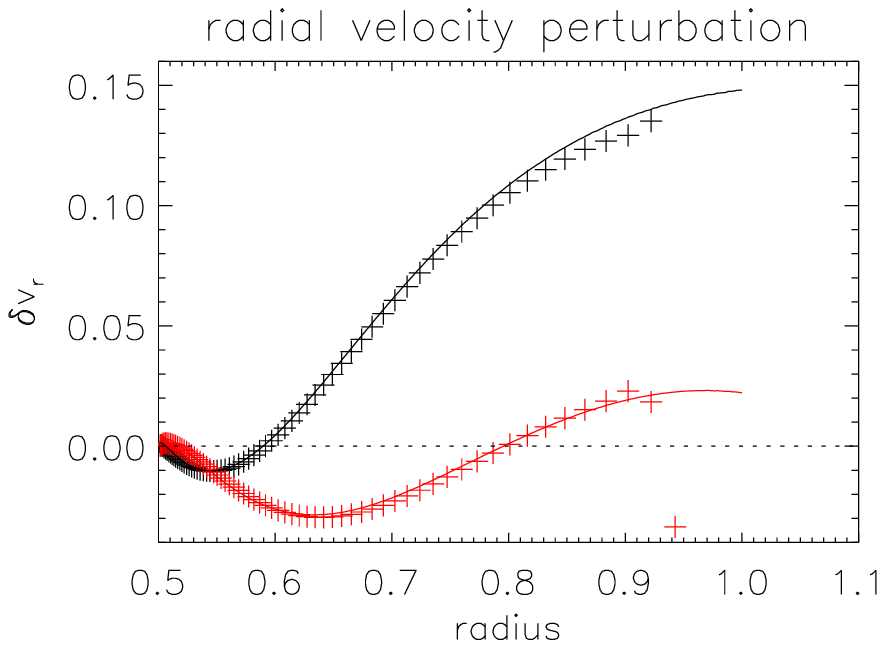}
  \includegraphics[width=\columnwidth]{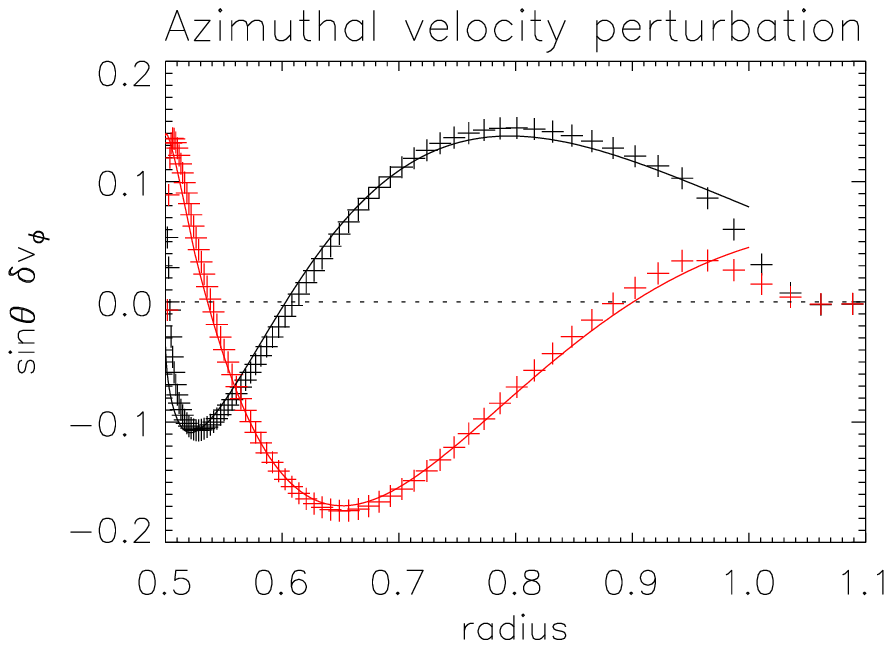}
 \caption{Radial profiles of radial velocity (left) and azimuthal
velocity multiplied by $\sin\theta$ (right), projected onto real $l=1$
spherical harmonics along the $x$-axis (black) and $y$-axis (red). The $+$ signs show values from model R5\_L11\_HR at $t=30$, while the full lines show the semi-analytical eigenmodes.} 
             \label{fig:perturbation_r5}%
\end{figure*}

\begin{figure*}
\centering
 \includegraphics[width=\columnwidth]{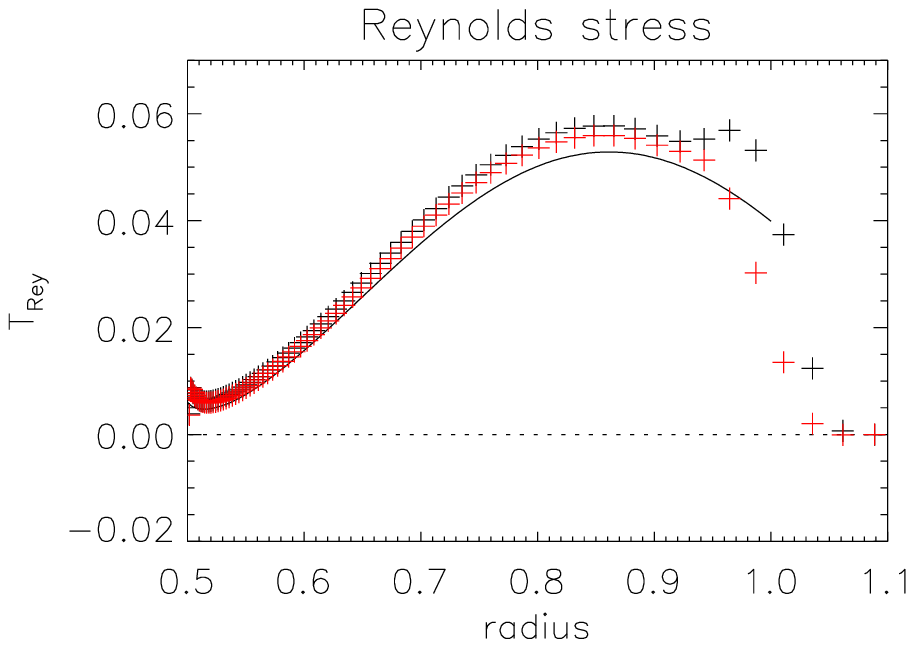}
  \includegraphics[width=\columnwidth]{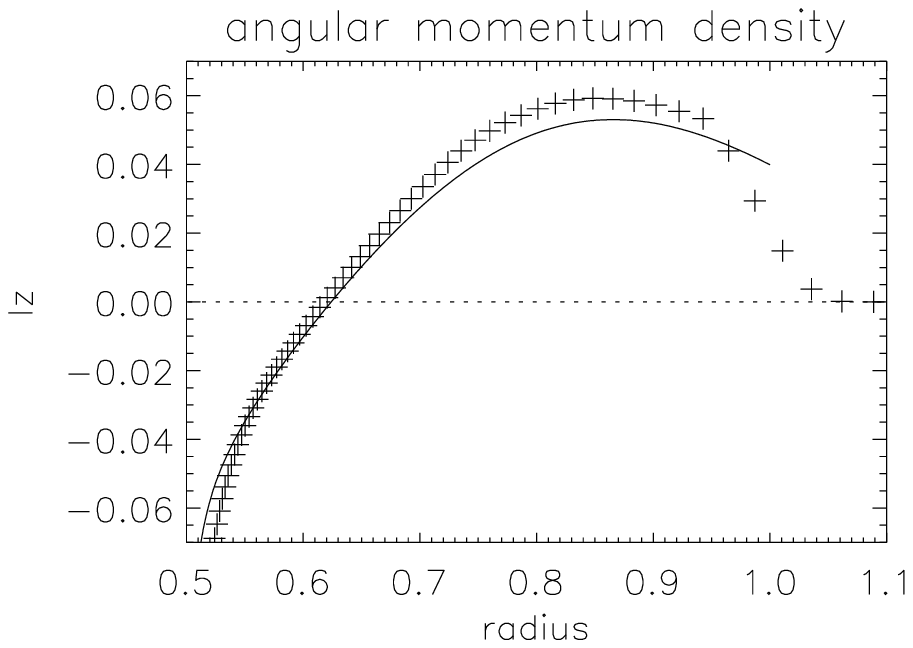}
 \caption{Radial profiles of the surface-integrated Reynolds stress (left) and 
          angular momentum density (right) resulting from an
          $l=1$ spiral mode with $r_*/r_{\rm s0}=0.5$. 
          The Reynolds stress is normalised by $\dot{M}r_{\sh0}|v_{\sh0}|$, while the angular momentum density is
          normalised by $\dot{M}r_{\sh0}$. The $+$ signs show values from 
          model R5\_L11\_HR at $t=30$, while the full lines show the semi-analytical predictions. The Reynolds stress is
          measured in the simulation in two different ways, as explained in the text (black and red colours). }
             \label{fig:lz_r5}%
\end{figure*}

Figure~\ref{fig:perturbation_r5} shows the projection of the radial velocity
and azimuthal velocity times $\sin\theta$ 
onto real spherical harmonics along the $x$ and $y$ axes. 
Profiles from the numerical simulation 
match those from linear theory with a very good accuracy 
showing that the method used to determine the amplitude of the spiral modes is
accurate.

Figure~\ref{fig:lz_r5} shows the radial profiles of the Reynolds stress and
angular momentum integrated over a spherical surface (defined by
Equations~(\ref{eq:def_Trey}) and (\ref{eq:def_lz}) respectively). The Reynolds
stress is extracted from the simulation in two different ways. First, we
subtract the advection of angular momentum by the mean flow ($v_rl_z$ where
$l_z$ is the surface integrated angular momentum density and $v_r$ is the
angular average of the radial velocity) from the surface integrated total
angular momentum flux defined by Equation~(\ref{eq:def_angmom_flux}) (the
result is shown in black). 
Alternatively, we compute the Reynolds stress caused by the
velocity perturbations projected on the $\{l,m\}=\{1,1\}$ 
spherical harmonic (red line).
These two measures are in good agreement (except in the vicinity of the shock) 
showing that the Reynolds stress is indeed dominated by the $\{l,m\}=\{1,1\}$ 
SASI mode, and that the second order expansion is valid (otherwise higher order terms would appear, see Section~\ref{sec:reynolds}).  

The analytically predicted shapes of the Reynolds stress and angular momentum profiles agree
very well with the numerical simulations. Radially inwards from the shock, the Reynolds stress first
slightly increases, and then smoothly decreases to a very small value near the
proto-neutron star surface. On the other hand, the angular momentum density
changes sign at an intermediate radius: angular momentum redistribution caused by the Reynolds stress creates a region
of positive angular momentum below the shock, and a region of negative
angular momentum above the PNS surface. 

\begin{figure*}
\centering
 \includegraphics[width=\columnwidth]{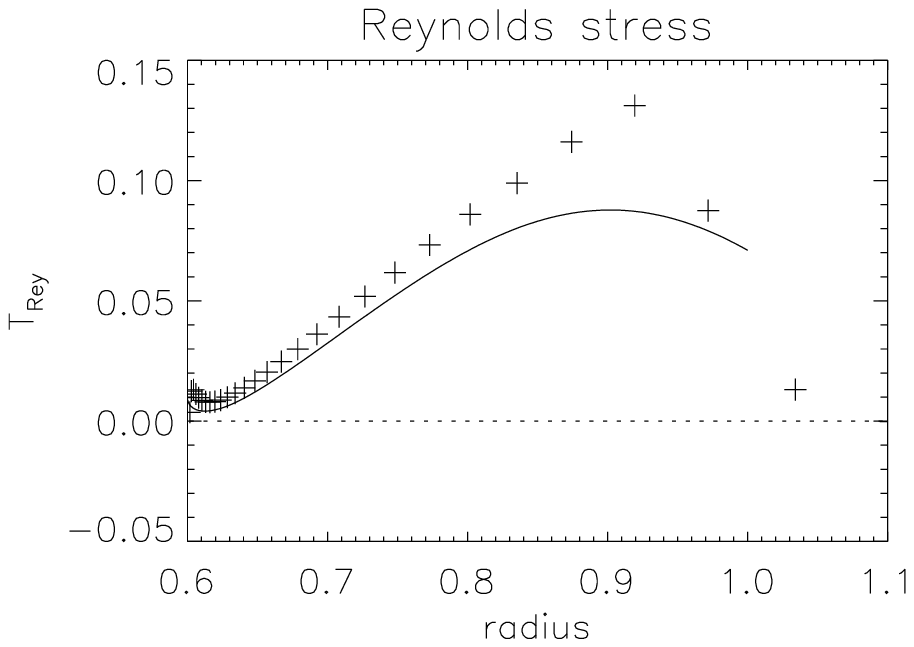}
  \includegraphics[width=\columnwidth]{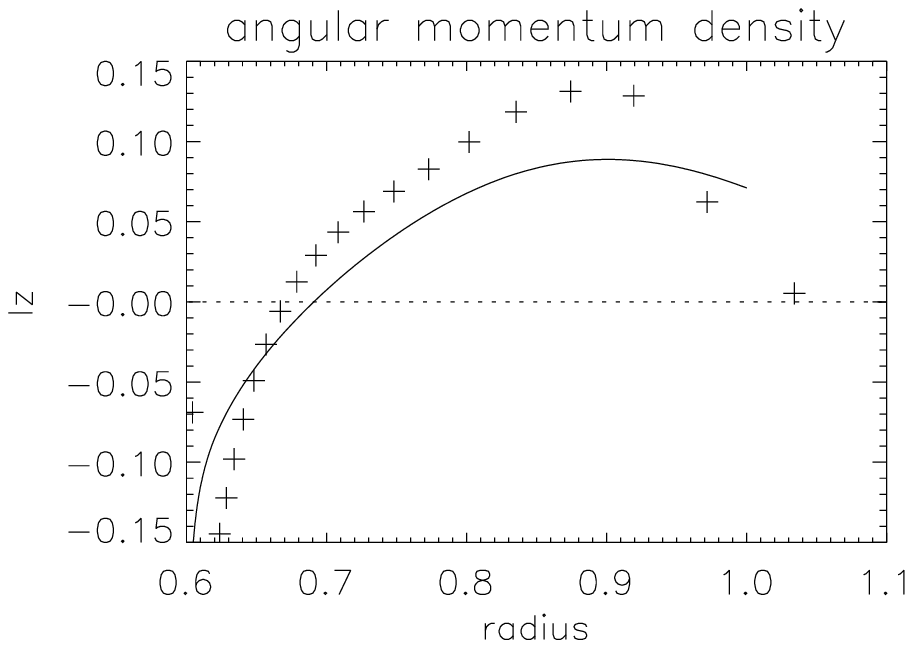}
 \caption{Radial profiles of the surface-integrated Reynolds stress (left) and angular momentum 
          density (right) resulting from a $\{l,m\}=\{2,2\}$ spiral with $r_*/r_\sh=0.6$.
          The $+$ signs show the result of model R6\_L22\_P2 
          at $t=30$, while the full lines show the semi-analytical predictions.}
             \label{fig:lz_r6}%
\end{figure*}

Note that the amplitudes of the Reynolds stress and angular momentum density are
higher in the numerical simulation by about $10-15\%$. The accuracy with which a
code describes the linear coupling at the shock has been studied by
\citet{sato09}, who argued that the relevant parameter is the ratio
$\lambda_{\rm adv}/\Delta r$, where $\lambda_{\rm adv}$ is the wavelength of
the advected wave created by the shock oscillation and $\Delta r$ is the radial
resolution. The radial resolution of the numerical simulation at the shock is
$\Delta r = 0.025r_\sh$. Using the frequency of the mode, we estimate
$\lambda_{\rm adv}/\Delta r = 50$. Comparing with Figure~10 of
\citet{sato09}, one may expect an accuracy of $\sim 20\%$, which is roughly
consistent with the accuracy we obtain when comparing the simulations with the
semi-analytical predictions.

%-------------------------------------------------
\subsection{Fundamental $l=2$ spiral mode}
	\label{sec:r6}

With a larger proto-neutron star $r_*/r_\sh=0.6$, modes with $l=1$ are stable 
and $l=2$ has only its fundamental mode destabilized. 
Depending on the initial perturbations, both $\{l,m\}=\{2,\pm1\}$ or $\{2,\pm2\}$ 
may be excited preferentially. We here focus on model 
R6\_L22\_P2, which is initially perturbed so that an 
$m=2$ spiral mode dominates. Using the same method as for $\ell=1$, 
we obtain the amplitudes of the two $m=\pm2$ spiral modes at time $t=30$: $A_2=0.185$ and
$A_{-2}=0.021$, respectively, showing that the $m=2$ mode is dominant and the $m=-2$ mode 
contributes at the $10\%$ level.

The Reynolds stress and angular momentum profile driven by this $\{l,m\}=\{2,2\}$
spiral mode are shown in Figure~\ref{fig:lz_r6}. Agreement between the numerical results and the analytical predictions 
is achieved within $30\%$, which is not as good as for $l=1$. This is most likely due to the twice lower resolution used in this simulation. 
Interestingly, the shapes of both of these profiles are very similar to those of the 
spiral $\{l,m\}=\{1,1\}$ mode of model R5\_L11\_HR. 
Note however that the amplitudes of the Reynolds stress and
of the angular momentum density for $\{l,m\}=\{2,2\}$ are larger 
by a factor of roughly two 
relative to that of the $l=1,m=1$ mode, despite the fact that the mode
amplitude is very similar. This factor is in fact predicted by 
equation~(\ref{eq:lzsh}), arising from the amplitude of the azimuthal velocity perturbation, 
which is proportional to the shock inclination in the azimuthal direction and therefore to the spherical
harmonic index $m$.

%-----------------------------------------------------------
\subsection{Higher frequency harmonics of $l=1$ spiral modes}
	\label{sec:r2}
	
\begin{figure*}
\centering
 \includegraphics[width=\columnwidth]{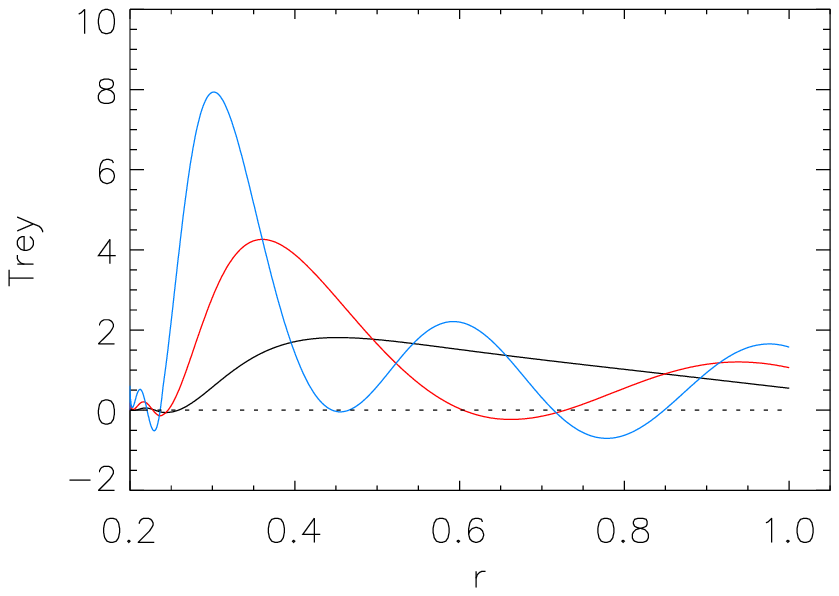}
 \includegraphics[width=\columnwidth]{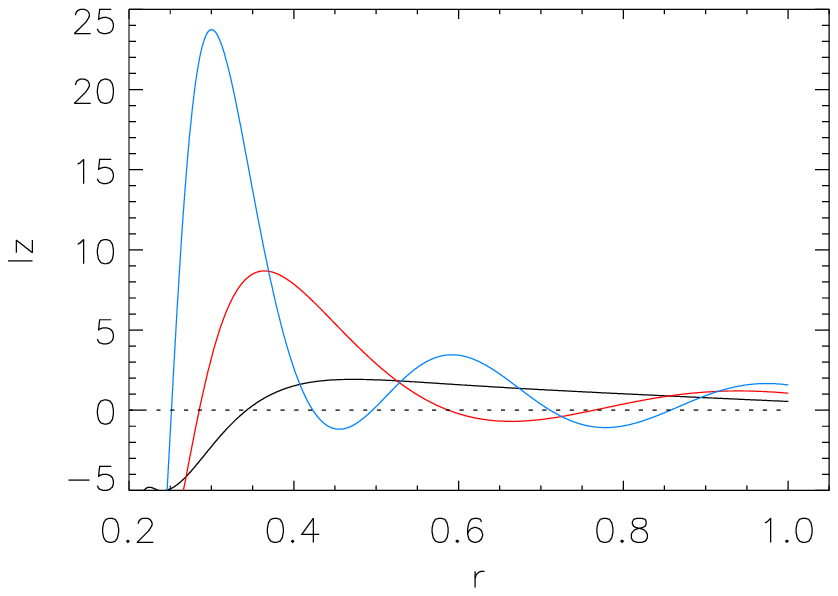}
 \caption{Analytical predictions for the surface-integrated 
          Reynolds stress (left) and angular momentum density (right) 
	  associated with the first three $l=1$ harmonics 
	  ($r_*/\rsh=0.2$). The fundamental mode is
          shown in black, while the first and second harmonics are represented by red and blue lines, respectively.}
             \label{fig:lz_harm}%
\end{figure*}

\begin{figure*}
\centering
 \includegraphics[width=\columnwidth]{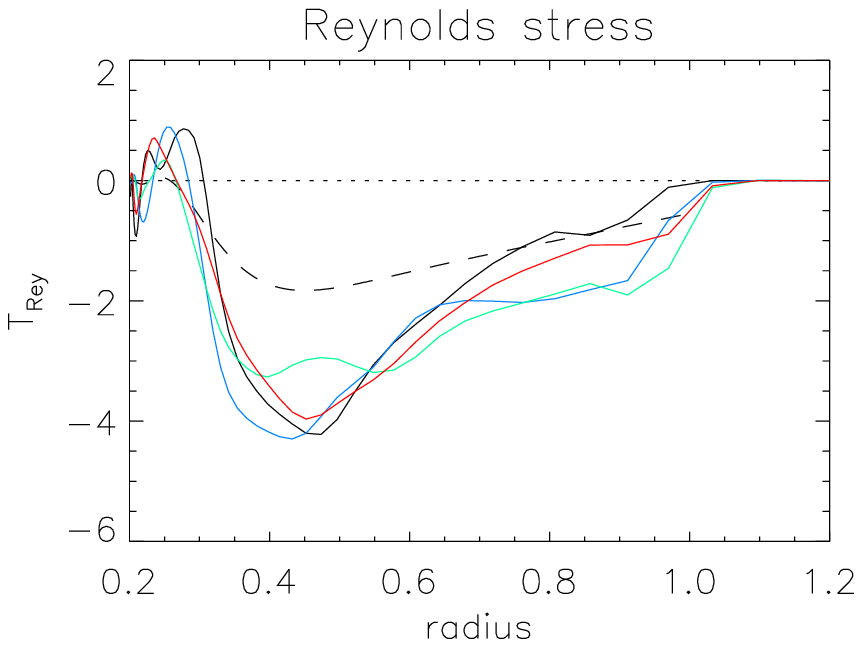}
 \includegraphics[width=\columnwidth]{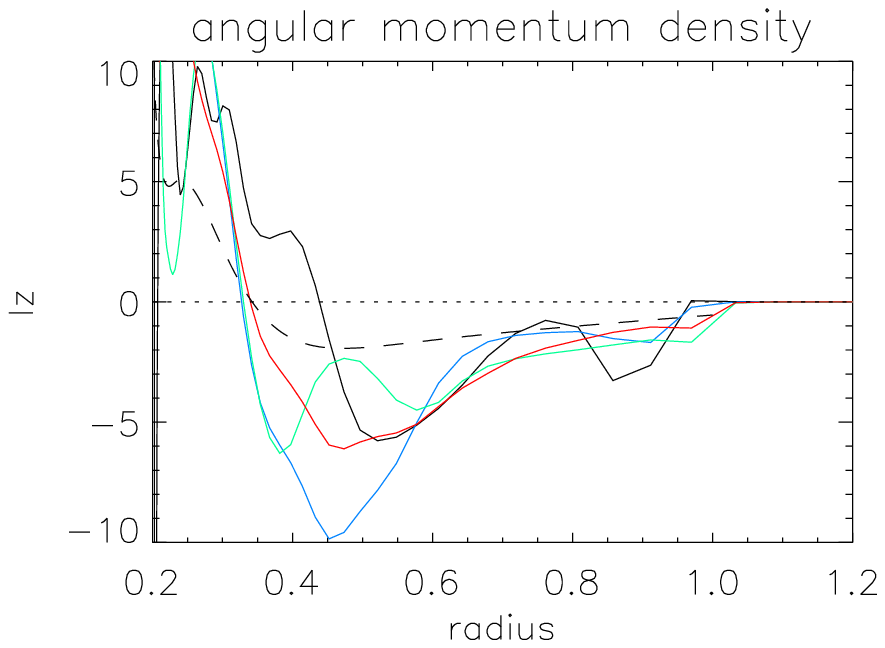}
 \caption{Surface-integrated Reynolds stress (left) and angular momentum density (right) 
in model R2\_L11h. The different colours represent
different times: $t=25$ (black), $t=30$ (blue), $t=35$ (green) and $t=40$
(red). These profiles are renormalised by the quantity $A_{-1}^2 - A_{1}^2$ (where $A_1$ and $A_{-1}$ are 
the amplitudes of the $m=1$ and $m=-1$
spiral modes, obtained by fitting the shock deformation projected onto $l=1$
harmonics). The Reynolds stress has been computed using the first method
described in Section~\ref{sec:r5}. Furthermore, the semi-analytical prediction
obtained with only the fundamental mode is shown with dashed lines.}
             \label{fig:lz_r2}%
\end{figure*}

For $r_*/\rsh=0.2$, several harmonics of $l=1$ 
are unstable. The frequencies of the fundamental mode, as well as the first and second harmonics are $\omega_r = \{0.53,1.02,1.52\}$, and their growth rates are $\omega_i = \{0.12,0.10,0.075\}$, respectively. The fundamental mode has the largest growth rate and should therefore play a dominant role, which is consistent with the simulations. However, the growth rate of the two higher frequency harmonics is only slightly smaller (specially for the first harmonic), suggesting that they could also have an impact on the dynamics and angular momentum redistribution. The individual contributions of these three modes to the surface-integrated
Reynolds stress and angular momentum density are illustrated in
Figure~\ref{fig:lz_harm}. As predicted by Equation~(\ref{eq:lzsh}), the angular
momentum density at the shock of a given mode is proportional to its frequency
and therefore increases with increasing harmonic order.

The Reynolds stress and angular momentum
density profiles generated by the fundamental mode are very similar to those 
obtained for different values of $r_*/\rsh$. 
In contrast, the profiles driven by the second and third
harmonics have a more complex structure exhibiting radial oscillations. The profiles exhibit one more oscillation per increasing harmonic order. 

These oscillations can be understood by decomposing the velocity perturbations below
the shock into waves as described in Appendix~\ref{sec:decomposition}. The
Reynolds stress is decomposed into six 
parts: three coming from the individual contributions of the vorticity wave and of
the two acoustic waves propagating up and down, and three additional
contributions arising from the interaction
between these waves. It is shown
in Appendix~\ref{sec:decomposition} that 
three of these contributions dominate:
the individual contributions from the vorticity wave and the 
acoustic wave propagating upwards both create a
\emph{non-oscillatory} Reynolds stress profile with
the same sign as $m$, while the interaction between these two waves drives a Reynolds stress that 
\emph{oscillates} in the radial direction between positive and negative values. The
three other contributions play only a very minor role. 
The analysis of Appendix~\ref{sec:decomposition} shows that 
the oscillatory contribution 
is more important in the higher frequency harmonics than in
the fundamental mode. Furthermore, if there is no phase shift in the wave coupling at
the shock and in the deceleration region close to the neutron star surface, 
then the analysis predicts one oscillation in the fundamental mode Reynolds stress
profile, two for the first harmonic, and three for the second harmonic. 
This is indeed the case in Figure~\ref{fig:lz_harm}, although the last half
oscillation is difficult to see because it lies very close the PNS surface. 

If there is a superposition of several $l=1$ unstable modes with different
frequencies, the resulting Reynolds stress is not merely a superposition of the
individual contributions from these modes. 
Instead, the interaction between the different modes
generates additional components that oscillate in 
time. One would therefore expect the shape of the
Reynolds stress profile to evolve in time 
and to display radial oscillations. 

Figure~~\ref{fig:lz_r2} shows the surface-integrated
Reynolds stress and angular momentum density 
in model R2\_L11h at times $t=25, 30, 35, 40$. These profiles are renormalised by dividing by
the quantity $A_{-1}^2 - A_{1}^2$ (where $A_1$ and $A_{-1}$ are the amplitudes
of the $m=1$ and $m=-1$ spiral modes obtained by fitting the shock deformation
projected onto $l=1$ harmonics). This renormalisation is chosen so that the
analytical prediction is the same 
if only the fundamental mode were present. 
While the overall time evolution of the
amplitude has been properly scaled out, the Reynolds stress and angular
momentum density profiles still show radial oscillations which are time dependent.
These oscillations might be attributed to the presence of smaller amplitude
higher frequency harmonic modes in addition to the dominant fundamental. 

Note that contrary to Section~\ref{sec:r5}, the Reynolds stress deduced from the simulation with the two different methods differ at small radii. This indicates that higher $l$ motions also play a role in determining the total Reynolds stress in this region.

%%%%%%%%%%%%%%%%%%%%%%%%%%%%%%%%%%%%%%%%%%%%%%%%%%%%%%%%%%%%
\section{Nonlinear phase}
	\label{sec:nonlinear}

All of the previous considerations are strictly valid only in the linear- or
weakly-nonlinear phase of the SASI. Several effects neglected in our analytical
treatment may appear in the fully non-linear phase: 
breakdown of mode linearity (saturation and higher-order terms), modification of `background' quantities, 
and  turbulent motions triggered by the SASI itself. Here we discuss the first and third effects in turn.

\subsection{Quasi-Steady-State Solution}
\label{sec:steady-state}

\begin{figure*}
\centering
  \includegraphics[width=\columnwidth]{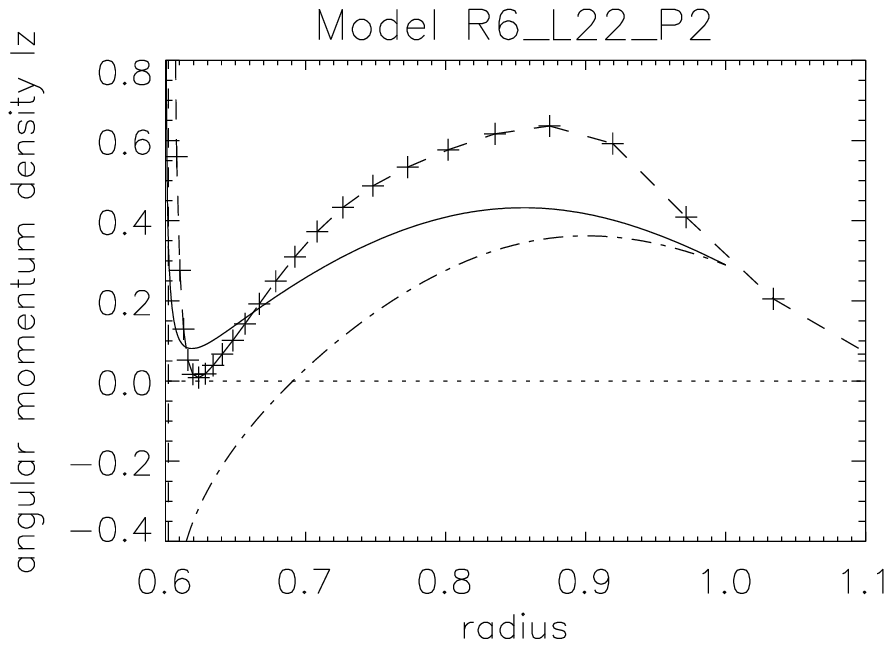}
  \includegraphics[width=\columnwidth]{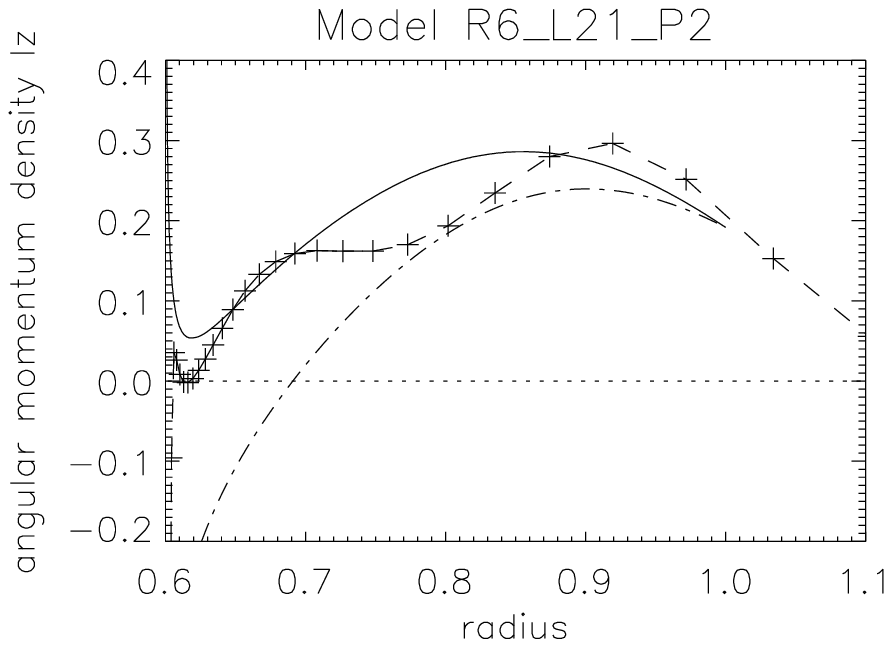}  
  
 \caption{Time-averaged profiles of surface integrated angular momentum density 
in the non-linear phase of models R6\_L22\_P2 ($\{l,m\}=\{2,2\}$, left panel) 
and R6\_L21\_P2 ($\{l,m\}=\{2,1\}$, right panel). Simulation results are shown with $+$ signs and a dashed line, while the analytical
predictions extrapolated to the saturated phase are shown with full black lines (from eq.~\ref{eq:lz_nonlinear}). For comparison, we also show
the predicted angular momentum density in the case of a growing mode 
(from eq.~\ref{eq:lzlm}, dot-dashed black line).}
             \label{fig:lz_r6_nonlinear}
\end{figure*}

In the non-linear phase, the amplitude of the SASI spiral mode
stops growing due in part to secondary instabilities 
(e.g., \citealt{guilet10b}). We can approximately take this into
account by assuming that the structure of the mode 
remains close to that given by the linear analysis, 
but then setting the growth rate to zero.
In this case, the evolution equation for the angular 
momentum (eq.~\ref{eq:lzlm_PDE}) becomes: 
\begin{equation}
\p_t l_{z} + \p_r(l_{z} v_0) = - \p_r T_{Rey},
	\label{eq:lz_evol_nonlinear}
\end{equation}
where $T_{Rey}$ is the Reynolds stress of the dominant spiral mode, which is now steady 
since the mode is not growing. This equation then admits the following stationary 
solution for the angular momentum profile:
\begin{equation}
l_{z} = -\frac{T_{Rey}}{v_0}.
	\label{eq:lz_nonlinear}
\end{equation}
This is the same as Equation~(\ref{eq:lzlm}) except for the absence of the second term due to
the growth of the mode. As a consequence, the predicted angular
momentum density has the same sign as the Reynolds stress, and therefore 
there is no sign change at an intermediate radius. 
This means that the matter with an angular momentum of opposite sign has been
accreted onto the proto-neutron and either accumulated in the very dense region
at the inner edge of the grid in the simulations, or left the numerical domain.

In Figure~\ref{fig:lz_r6_nonlinear}, the solution to equation~\ref{eq:lz_nonlinear} is compared with time-averaged profiles from models 
R6\_L22\_P2 (dominated by an $\{l,m\}=\{2,2\}$ spiral mode)
and R6\_L21\_P2 (dominated by an $\{l,m\}=\{2,1\}$ mode). The time average is performed over the time
interval $t=[50,100]$, during which the spiral modes have an approximately
constant amplitude. This amplitude can be measured by performing a fit like in
Section~\ref{sec:linear}, because the time evolution of the shock deformation
is close to a sinusoid. The resulting amplitude is then used to normalize
the analytic prediction.

Although the assumptions made in the analytical treatment are not well justified in the non-linear phase, Figure~\ref{fig:lz_r6_nonlinear} shows that the
analytical predictions are in fairly good agreement with the time-averaged profiles of angular momentum density. 
In particular, the angular
momentum density profile does not change sign except very close to $r_*$, in contrast
to the linear phase, being
much better predicted by Equation~(\ref{eq:lz_nonlinear}) (full line) 
than by Equation~(\ref{eq:lzlm}) (dotted line). 
This rather good agreement is helped by the fact that the saturation amplitude 
is fairly low (compared to the simulation with $l=1$ SASI activity), and therefore the shock does not
significantly expand due to the spiral mode activity.

\begin{figure}
\centering
  \includegraphics[width=\columnwidth]{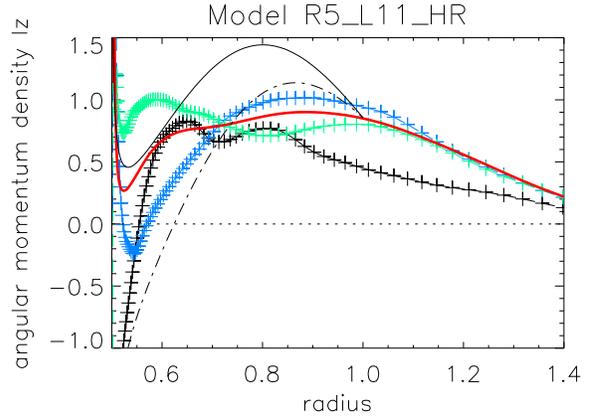}
 \caption{Surface-integrated angular momentum density in the non-linear phase of 
model R5\_L11\_HR. The $+$ signs with dashed lines show instantaneous profiles
at times $t=50$ (beginning of the saturated phase, black line), $t=65$ (blue line), and
$t=75$ (green line). The red full line shows the time-averaged profile in the time
interval $t=[75,125]$, and the black full line shows the analytical prediction extrapolated to the saturated phase
(using eq.~[\ref{eq:lz_nonlinear}]). For comparison, we also show with a dot-dashed
black line the predicted angular momentum density in the case of a growing mode
computed using Equation~(\ref{eq:lzlm}).}
             \label{fig:lz_r5_nonlinear}%
\end{figure}

Figure~\ref{fig:lz_r5_nonlinear} shows results from the non-linear phase of 
model R5\_L11\_HR, which is dominated by a large amplitude 
$\{l,m\}=\{1,1\}$ spiral mode. 
In this model as well, the time-averaged angular momentum
density profile does not change sign except very close to the PNS surface. 
In order to illustrate the time-evolution
towards this new shape of the surface-integrated angular momentum density, 
we plot the instantaneous profiles at three different times: $t=50$, 
just before the amplitude of the SASI mode saturates, 
and two later times ($t=65$ and $t=75$) that are well into the non-linear saturated 
phase. There is a large variation in the shape of the radial profile, 
most notably the radius where the angular momentum changes sign moves inward.

In order to compare the time-averaged profiles with the semi-analytical
results, we estimate the amplitude of the spiral mode in the saturated state
as
\footnote{We did not measure the spiral mode amplitude with 
the method of Appendix~\ref{sec:spiral_mode_amplitude}
because the time evolution of the shock deformation projected onto
spherical harmonics is irregular, yielding a poor-quality fit. We note that this 
alternative method to measure the spiral mode amplitude is accurate if a
single spiral mode dominates the dynamics. For models R6\_L22\_P2 and
R6\_L21\_P2, where a fit could be performed showing a 
dominant single spiral, the alternative method gives the same results as the fit.}
\begin{equation}
A_1 = \sqrt{2}\left(\overline{a_x^2 + a_y^2}\right)^{1/2},
	\label{eq:spiral_amplitude_nonlinear}
\end{equation}
where $a_x$ and $a_y$ are the shock deformation amplitude projected onto real
spherical harmonics along the $x$ and $y$ axis, the bar represents a time average,
and the factor $\sqrt{2}$ accounts for the 
different normalisation of real and complex
spherical harmonics. For model R5\_L11\_HR we obtain an 
amplitude of $A_1 = 1.03$. Due to this large amplitude spiral mode, 
the shock significantly expands compared to its initial position, 
reaching an average radius of $\simeq 1.4r_{\sh0}$. As a result, 
the time-averaged profiles of angular momentum density differ from the analytical predictions in that it extends to larger radii. It is also flatter
than the analytical prediction, with a value close to the analytical value predicted at the shock.

%----------------------------------------------------------------------------
\subsection{Reynolds decomposition}
\label{sec:reynolds}

To gain further insight into the angular momentum redistribution
in the non-linear phase,
we perform a Reynolds decomposition on the models 
of Table~\ref{tab:lz_tot} and make use of conservation laws.
Hereafter, the symbol $\langle A \rangle$ is used to denote the time- and 
angle-average of a quantity $A$, 
\begin{equation}
\label{eq:average_definition}
\langle A \rangle (r)\equiv \frac{1}{4\pi(t_f - t_i)}\int_{t_i}^{t_f}\dd t\,
		       \int_{4\pi} \dd \Omega\, A(r,\theta,\phi,t),
\end{equation}
where $t_i$ and $t_f$ are the initial and final times chosen for the
averaging interval.

In a saturated state that is stationary in a time averaged sense, 
conservation of angular momentum (equation~\ref{eq:ang_mom_conservation}) becomes
\begin{equation}
\p_r\,\langle \F \rangle =  4\pi\p_r\,\langle r^2 \rho v_r\, (r\sin\theta v_\phi)\rangle = 0.
	\label{eq:ang_mom_conservation_sat}
\end{equation}
In the absence of rotation, we have $\langle\F\rangle=0$ upstream of the shock, 
therefore $\langle\F\rangle=0$ should be verified everywhere in the flow. 
Similarly, the equation of time-averaged mass conservation reads
\begin{equation}
\p_r\,\langle r^2\rho v_r\rangle = 0.
	\label{eq:mass_conservation_sat}
\end{equation}
The mass accretion rate above the shock then sets the value of the
mass flux everywhere, 
$\langle r^2 \rho v_r\rangle= -\dot{M}/4\pi$. 

Now let us decompose the density, radial velocity,
and specific angular momentum 
\begin{equation}
\lambda\equiv r\sin\theta v_\phi 
\end{equation}
into a mean value plus a fluctuating component with vanishing average,
\begin{eqnarray}
\rho    & = & \langle \rho\rangle + \Delta \rho, \\
v_r     & = & \langle v_{r}\rangle + \Delta v_r, \\
\lambda & = & \langle \lambda\rangle + \Delta \lambda,
\end{eqnarray}
Note that contrary to previous sections, the fluctuating component is not assumed to be small. 
We will focus the discussion on the time-averaged, surface-integrated 
angular momentum flux,
\begin{equation}
\label{eq:ang_flux_ave_def}
\langle\F\rangle = 4\pi r^2\, \langle\rho v_r \lambda\rangle.
\end{equation}

Separating the radial velocity and specific angular momentum into mean and
fluctuating components, we obtain
\begin{equation}
\label{eq:ang_flux_ave_terms}
\langle \F \rangle = 4\pi r^2 \left[\langle \rho \lambda\rangle \langle v_r\rangle  +
		                     \langle \rho \Delta v_r \Delta \lambda\rangle +
	  	                     \langle \Delta \rho \Delta v_r\rangle \langle \lambda\rangle\right].
\end{equation}
We recognise the first term on the right hand side as the angular momentum density advected
by the mean flow, the second term as the mean Reynolds stress 
\begin{equation}
\label{eq:reynolds_reynolds}
\langle T_{Rey}\rangle \equiv 4\pi r^2\,\langle \rho \Delta v_r \Delta \lambda\rangle,
\end{equation}
and the third term as the angular momentum transported by the fluctuating component of the mass flux (see, e.g., 
\citealt{murphy11} for the physical meaning of terms in the Reynolds-averaged fluid equations).

From equation~(\ref{eq:ang_flux_ave_terms}) we can solve for the mean angular momentum density
\begin{equation}
\langle l_z\rangle =  - \frac{\langle T_{Rey}\rangle}{\langle v_r\rangle} 
		       -\frac{4\pi r^2 \langle\lambda\rangle}{\langle v_r \rangle }\langle \Delta\rho \Delta v_r\rangle
                       +\frac{\langle \F\rangle}{\langle v_r \rangle}.
	\label{eq:lz_sat1}
\end{equation}
The first term of on the r.h.s. mirrors the corresponding term in
equations~(\ref{eq:lzlm}) and (\ref{eq:lz_nonlinear}). 
The second term is not present in equation~(\ref{eq:lzlm}) because
it is of higher order in the expansion of small mode amplitude. 

\begin{figure}
\includegraphics*[width=\columnwidth]{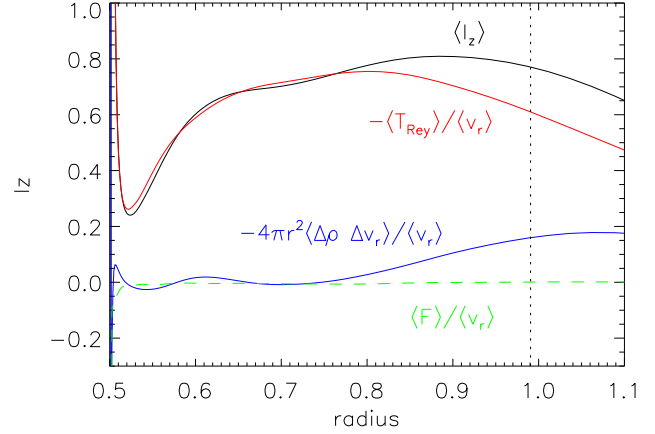}
\caption{Time-averaged and surface-integrated angular momentum density as a function of radius
(equation~\ref{eq:lz_sat1}) for model R5\_L11\_HR (black). Also
shown are the contributions from the mean Reynolds stress (red), the fluctuating component
of the mass
flux (blue), and the total angular momentum flux (green). 
The vertical dashed line 
marks the innermost radius affected by shock oscillations, defined
as the average minimum shock radius minus one standard deviation \citep{fernandez09a}.}
\label{f:angz_ave_HR}
\end{figure}

Figure~\ref{f:angz_ave_HR} shows the different terms that 
make up equation~(\ref{eq:lz_sat1}) applied to model R5\_L11\_HR, with
the time-average taken over the interval $[75,125]t_{\rm ff0}$.
The angular momentum flux $\langle \F\rangle$ is smaller than the
other terms by 2 orders of magnitude over most of the domain, verifying that global angular
momentum conservation is satisfied to a reasonable degree. Very close to the 
neutron star surface, matter piles up and the system is never in steady-state, hence
the non-zero $\langle \F \rangle$ in that region.

The dominant contribution to the angular momentum density arises from the
Reynolds stress, justifying equation~(\ref{eq:lz_nonlinear}) a posteriori.
Near the shock, the component due to the fluctuating mass flux becomes
important, though it never exceeds that due to the Reynolds stress. Note that
in order to maintain steady-state, both terms must transport angular momentum
outward to counteract advection by the mean flow.

The Reynolds stress has contributions from the large-scale saturated spiral 
mode as well as from smaller scale
turbulent fluctuations. Further analysis in the space-time frequency domain could 
separate these components; we leave this for future work.

%%%%%%%%%%%%%%%%%%%%%%%%%%%%%%%%%%%%%%%%%%%%%%%%%%%%%%%%%%%%%%%%%%%
\section{Approximate expression for the maximum angular momentum 
contained in a spiral wave}
	\label{sec:approximate}
	
A SASI spiral mode separates the postshock flow into a 
region below the shock where matter rotates
in the same direction as the spiral mode, and another region further below where 
angular momentum has the opposite sign. The magnitude of this redistribution is
\begin{equation}
L_z = \int_{r_0}^{\rsh} l_z\,\dd r,
\end{equation}
where $r_0$ is the radius where the angular momentum density changes sign.  We
found in Section~\ref{sec:nonlinear} that during the saturated phase, 
the angular momentum density profile in the SASI active region does
not change sign except very close to the protoneutron star, indicating that 
matter with the opposite sign of angular momentum has been already accreted. 
We will therefore approximate the radius $r_0$ by the radius of the proto-neutron star
$r_*$. Figure~\ref{f:angz_ave_HR} suggests that the angular momentum density has
a rather flat profile; we will thus assume that the angular
momentum density equals its value below the shock everywhere in the postshock
region. Note that this assumption is only approximate and is based on an empirical
observation rather than a strong theoretical argument. It would therefore be
useful to check its validity in numerical simulations of less idealised flows,
in particular including neutrino heating. The results of Section~\ref{sec:r2}
and Appendix~\ref{sec:decomposition} show that the fundamental mode creates an
angular momentum profile with much less radial structure than higher frequency
harmonics. This suggests that our assumption might remain approximately valid
as long as the dynamics is dominated by the fundamental mode. 

Using these two simplifying assumptions, we can write the angular momentum
magnitude as
\begin{equation}
L_z  \simeq (\rsh-r_0)l_{z\sh} \simeq (\rsh-r_*)l_{z\sh}.
	\label{eq:Lz0}
\end{equation}
This equation can then be combined with equation~(\ref{eq:lzsh}) to obtain an
analytical estimate of the total angular momentum redistributed by a SASI
spiral mode 
\begin{equation}
L_{z} \simeq m f(\kappa,\M_1) \frac{\omega_r(\rsh-r_*)}{2\pi |v_\sh|}\dot{M}\rsh^2 \left(\frac{\Delta r}{\rsh}\right)^{2}.
 	\label{eq:Lz_tot}
\end{equation}
Note that equation~ ~(\ref{eq:lzsh}) is still valid in the presence of neutrino heating, which (just like neutrino cooling) affects the angular momentum density below the shock only indirectly  through its effect on the frequency of the mode and its saturation amplitude. For the fundamental mode, the frequency is approximately $\omega_r \sim
2\pi/\tau_{\rm aac}$, where $\tau_{\rm aac}$ is the advective acoustic time
\citep{foglizzo07,guilet12}, thus
\begin{equation}
\frac{\omega_r(\rsh-r_*)}{2\pi |v_\sh|} \simeq \frac{\tau_{\rm aac}(\rsh-r_*)}{|v_\sh|}.
\end{equation}
This ratio is expected to be $\lesssim 1$, because  the advective-acoustic time $\tau_{\rm aac}$
is slightly longer than the advection time from the shock to the neutron star $\tau_{\rm adv}$, 
and the advection time in turn is longer than that estimated with a constant velocity, $(\rsh-r_*)/|v_\sh|$,  
because the flow is decelerated.
The flows studied in Section~\ref{sec:linear} with different values of
$r_*/r_\sh$ all satisfy $\omega_r(\rsh-r_*)/(2\pi |v_\sh|) \simeq 0.4$ (see
Table~\ref{tab:lz_tot}). 

In Table~\ref{tab:lz_tot}, we compare the angular momentum redistributed by a
SASI spiral mode predicted by Equation~(\ref{eq:Lz_tot}) with the results of
the numerical simulations of \citet{fernandez10}. For this purpose the spiral
mode amplitude is measured in the simulations following
equation~(\ref{eq:spiral_amplitude_nonlinear}), the spiral mode frequency is
taken to be that predicted by the linear analysis, and the shock radius that of
the initial stationary state. 
For most simulations, the analytical prediction differs from the numerical result
by a few tens of percents which is comparable to the accuracy found in the
previous sections (the largest difference is for simulation R6\_L22\_P2 
at $45\%$). Note that the analytical prediction tends to slightly
underestimate the angular momentum measured in the numerical simulations; 
this might be due partly to the fact that the mean shock radius has expanded, 
allowing more angular momentum to accumulate. 

We now turn to discuss the significance of these results for the spin of
neutron stars at birth. The saturated SASI phase should end at the onset of explosion.
The magnitude of the angular momentum imparted to the neutron star will depend
on the radius separating expelled and accreted matter: if it is too deep in the postshock region in the case of an early explosion, 
matter with both signs
of angular momentum will be ejected, reducing the total angular momentum
\citep{rantsiou11}. Also, the magnitude of the angular momentum 
is also reduced if SASI activity occurs episodically as observed by \citet{hanke13} and \citet{iwakami13}, in
which case angular momentum with alternating signs is  accreted onto the
proto-neutron star. Nevertheless, we can estimate the maximum imparted angular momentum
in the idealised scenario outlined by \citet{blondin07a}, where all the matter rotating in the same
direction as the SASI spiral mode is ejected.

\begin{figure}
\includegraphics*[width=\columnwidth]{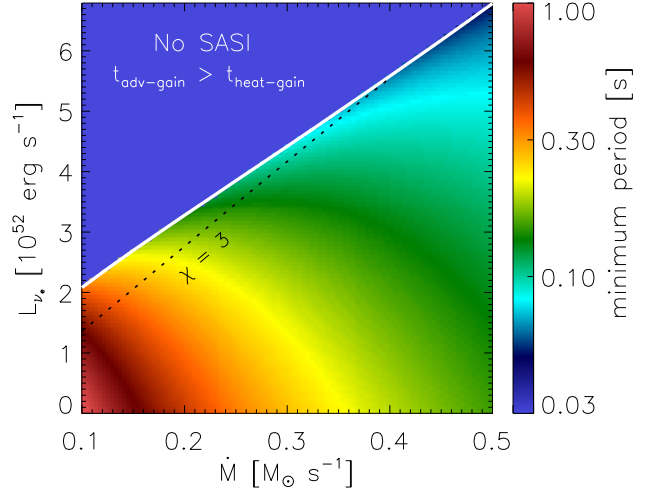}
\caption{Minimum neutron star rotation period that can be generated
via a spiral SASI mode, as inferred 
from equation~(\ref{eq:minimum_period}). The input parameters (shock radius,
shock compression ratio, SASI period, and postshock velocity) are computed from the steady-state solutions
of \citet{fernandez12}, which employ a realistic equation of state. The stellar radius is set to $r_* = 30$~km, and
the steady-state shock radius has been
multiplied by a factor $(1+\Delta r / r_{\rm s0})$, with $\Delta r / r_{\rm s0} = 0.3$ a typical saturation
value of the SASI. The region marked `No SASI' is such that the runaway condition in spherical symmetry
\citep{janka98,thompson00,thompson05} is met, 
thus the SASI does not have time to develop before explosion. 
The dashed
line shows the threshold $\chi=3$, below which the SASI is 
expected to dominate the dynamics \citep{foglizzo06}.}
\label{f:minimum_period_2d}
\end{figure}

Evaluating Equation~(\ref{eq:Lz_tot}), we obtain 
\begin{eqnarray}
L_{z} &\simeq& 2.3\times 10^{46} \left(\frac{\kappa}{10}\right)\left(\frac{50\,{\rm ms}}{P_{sasi}}\right)\left(\frac{\rsh-r_*}{120\,{\rm km}}\right)\left(\frac{3000\,{\rm km.s^{-1}}}{v_\sh}\right) \nonumber \\
&& \times \left(\frac{\dot{M}}{0.3\,{\rm M_\odot.s^{-1}}}\right)\left(\frac{\rsh}{150\,{\rm km}}\right)^2 \left(3\frac{\Delta r}{\rsh}\right)^{2} \,{\rm g.cm^2.s^{-1}}.
 	\label{eq:Lz}
\end{eqnarray}
Assuming a moment of inertia of the neutron star of $I = I_{45} \times
10^{45}\,{\rm g.cm^2}$,  this can be translated into a minimum period of
uniform rotation 
\begin{eqnarray}
P &\simeq& 290 \, I_{45}\left(\frac{10}{\kappa}\right)\left(\frac{P_{sasi}}{50\,{\rm ms}}\right)\left(\frac{120\,{\rm km}}{\rsh-r_*}\right)\left(\frac{v_\sh}{3000\,{\rm km.s^{-1}}}\right) \nonumber \\
&& \left(\frac{0.3\,{\rm M_\odot.s^{-1}}}{\dot{M}}\right)\left(\frac{150\,{\rm km}}{\rsh}\right)^2 \left(\frac{\rsh}{3\Delta r}\right)^{2} \,{\rm ms}.
 	\label{eq:minimum_period}
\end{eqnarray}
Note the dependence on the square of the amplitude and the shock radius.

We have so far applied our analytical results to an idealised setup where only SASI develops due to the absence of heating. Our analytical treatment can in principle be applied more generally as long as a SASI spiral mode dominates the dynamics, and a comparison with more realistic simulations would be desirable. Figure~\ref{f:minimum_period_2d} shows the result of evaluating
equation~(\ref{eq:minimum_period})
with parameters from the steady-state accretion shock models of \citet{fernandez12}. These
solutions employ the equation of state of \citet{shen98} as implemented by \citet{oconnor10},
and use a `lightbulb' approximation to neutrino heating.
The minimum period is computed as a function of the mass accretion rate and electron
neutrino luminosity, taking\footnote{Other parameters are the same as in the $R_\nu=30$~km
sequence of \citet{fernandez12}.} $r_* = 30$~km. 
To account for the fact that in the saturated SASI phase
the average shock radius is
larger than the initial steady-state value, 
we multiply $r_{\rm shock}$ in equation~(\ref{eq:minimum_period})
by $(1 + \Delta r / r_{\rm s0})$ and set $\Delta r / r_{\rm s0} = 0.3$. 
Note however that we are ignoring the effects of convection on this 
parameterization of the saturated SASI amplitude, basing it instead on
results from simulations without neutrino heating. Our analytical treatment also assumes a constant dissociation energy at the shock, which is only an approximate description with the equation of state employed here.

The trend of Figure~\ref{f:minimum_period_2d} is evident: shorter periods are obtained 
with larger neutrino luminosities -- which yield larger shock radii -- and larger accretion rates.
The normalization indicates that massive progenitors with large accretion rates, where strong SASI activity is expected
\citep{mueller12,hanke13,ott2013,iwakami13}, can lead to periods $\sim 100$~ms or less. In contrast, progenitors
that have a lower accretion rate and which may be expected to suppress SASI activity 
(e.g., \citealt{mueller12,takiwaki2012,murphy2012,dolence13,couch13b}),
would otherwise acquire very moderate amounts of angular momentum if the SASI were present, 
with minimum periods in the range $0.3-1$~s.  Note that the latter value is comparable to the spin periods obtained by \citet{wongwathanarat10,wongwathanarat13}.

From an observational point of view, the spin of neutron stars at birth is
still poorly constrained. The difficulty comes from the fact that the observed period of pulsars is very different from their initial period because of spin
down, and that the true age of most pulsars is unknown. Population synthesis studies nevertheless suggest that a distribution of
initial spin peaking around $300$~ms is consistent with observations (e.g.,
\citealt{F-G2006}). 
The age of some pulsars can be estimated when they are
associated with a supernova remnant, which then allows to constrain their
initial spin period. 
Despite poor statistics and sometimes large uncertainties,
these observations suggest that a significant fraction of neutron stars have
initial periods longer than $100$~ms (e.g. \citealt{popov12} and references
therein). The range of pulsar spin periods we obtain is therefore of the same
order of magnitude as that inferred from the observations, and we conclude that
angular momentum redistribution by a SASI spiral mode can be relevant to explain these observations. Note, however, that rotation initially present in the progenitor, which was neglected in this study, could also add a significant contribution to the angular momentum of the neutron star. 

%%%%%%%%%%%%%%%%%%%%%%%%%%%%%%%%%%%%%%%%%%%%%%%%%%%%%%%%%%%%%%%%%%%%%%%%%%%%%%%%%%%%%
\section{Conclusions}
	\label{sec:conclusion}
	
We have developed an analytical description of the angular momentum
redistribution driven by SASI spiral modes. It is based on a second order
perturbative expansion of the flow, which is valid when the amplitude of the spiral mode is
small. 

Angular momentum redistribution is due to the Reynolds stress of the
SASI mode, which can be computed using a linear analysis (\S\ref{sec:formalism}). 
For the lowest frequency SASI modes, this Reynolds stress has the same sign as the spherical
harmonic index $m$ of the mode. This causes angular momentum with the same
rotation direction as the spiral mode to accumulate below the shock, while
angular momentum with the opposite sign is accreted onto the proto-neutron
star. 

Higher frequency harmonics have more complex Reynolds stress profiles,
showing radial oscillations (Fig.~\ref{fig:lz_harm}). These can be explained by 
decomposing the velocity perturbations into contributions from the vorticity wave 
and two acoustic waves
propagating up and down (Appendix~\ref{sec:decomposition}). 
The individual contributions of the vorticity wave and
the acoustic wave propagating up create a non-oscillating Reynolds stress
profile with the same sign as $m$, while the interaction between these two
waves causes radial oscillations of the Reynolds stress, with more oscillations
being present for the higher frequency harmonics. In a realistic core-collapse
supernova context, where multiple modes can be excited by either initial
perturbations or convection, a situation with multiple unstable modes is
more likely to be obtained.

These analytical results compare favorably with the 3D
simulations of \citet{fernandez10}, the Reynolds stress and angular momentum
profiles in the linear phase agreeing within $10\%$ for the best resolved simulation, and a few tens of percent at lower resolution.

Although strictly speaking the analytical results
are valid only in the linear phase, we have found that they give a reasonable description of the angular momentum density
in the non-linear phase if the second term of Equation~(\ref{eq:lzlm}) is omitted to account for the fact that the spiral mode is not growing anymore.
As a consequence, nearly all of the SASI active region has the same sign of angular momentum because 
the matter with opposite angular momentum has already been accreted. 

We have also performed a Reynolds
decomposition of the numerical models of \citet{fernandez10} 
in the saturated phase. The Reynolds stress is again the dominant agent determining the
angular momentum profile below the shock. In addition, a contribution
from the fluctuating mass flux becomes important near the shock. Both
of these effects transport angular momentum outwards, balancing inward transport by advection.

Finally we derived an approximate analytical expression for the angular
momentum contained in the SASI spiral wave (eq.~\ref{eq:Lz_tot}). 
This expression depends on the mass accretion
rate, the shock and PNS radius, the compression ratio at the shock, and the
characteristics of the SASI spiral mode (frequency and amplitude). This allows
us to estimate the maximum angular momentum that can be imparted to the neutron
star if all the SASI active region is ejected during the explosion. The expected minimum neutron star spin periods in uniform rotation (Fig.~\ref{f:minimum_period_2d}) are consistent with values estimated by observations of pulsars associated with supernova remnants and by population synthesis studies for the bulk of the pulsar population.
Our analysis further suggests that the angular
momentum of the nascent neutron star should be positively correlated with the
mass accretion rate at the time of explosion if progenitors are slowly rotating. 
As a consequence, neutron stars born from progenitors with a shallow density profile 
-- for which the SASI should dominate the explosion dynamics (e.g., \citealt{mueller12,hanke13,iwakami13}) -- 
should rotate faster on average 
than those arising from stars with steeper profiles, which are generally less massive.

Obviously, the above prediction is contingent on a very idealised scenario in 
which the mass cut at explosion coincides with the surface where the angular momentum 
changes sign. The results of \citet{hanke13} indicate that even in progenitors where
strong SASI activity is expected, an episodic occurrence of spiral modes can result in
no net angular momentum being imparted to the neutron star. Prolonged SASI activity
up to the point of explosion, as seen in the 2D models of \citet{mueller12}, is essential
for this spin-up mechanism to work. It also requires at least one spiral mode (with spherical hamonics $\{l,m\}$) to dominate over the counterrotating spiral mode (with spherical harmonics $\{l,-m\}$). This may arise naturally from a symmetry breaking that has been observed in some numerical simulations and in the SWASI experiment \citep{blondin07a,fernandez10,foglizzo12}, but the timescale and conditions in which this occurs still need to be better understood.

Our analysis applies if the initial rotation of the progenitor is
negligibly slow. More generally, the initial spin of neutron stars 
is likely to result from a combination of angular momentum initially present in the
progenitor and that redistributed by the SASI. Given that 
prograde modes are expected to grow faster \citep{yamasaki08}, the dominance of one such mode would impart angular momentum to the neutron star with a sign opposite to that of the progenitor, in the idealised scenario in which the explosion carries away all the angular momentum of a given sign \citep{blondin07a}. Further studies using a rotating progenitor will be needed to clarify the consequences on the dynamics and on the spin of neutron stars.

We emphasize that the analytical formula for the angular momentum redistributed
by a spiral mode depends strongly on the amplitude of the spiral mode. In this study, we did not try to determine
analytically the amplitude of the spiral mode, and simply took it as an input
from the simulations. A semi-analytical description of the saturation of SASI
has been obtained by \citet{guilet10b}. They studied the ability of parasitic
instabilities (of Rayleigh-Taylor or Kelvin-Helmholtz type) to grow on a SASI
mode and destroy its coherence when it reaches a critical amplitude, thus
leading to its saturation. A remaining theoretical uncertainty is the role
of shock kinks. Future work addressing the 
saturation of the SASI can benefit from our theoretical estimate of the 
maximum angular momentum in the system.

Finally, we note that the present study neglects the effects of magnetic
fields, which can transport angular momentum via the Maxwell stress. The influence of a magnetic field on the linear growth of SASI has been studied in a planar toy model by \citet{guilet10a}. One possible extension of the present study is including magnetic
effects in the angular momentum redistribution in spherical or even
cylindrical coordinates. \citet{guilet11} have shown that Alfv\'en waves can be amplified in the
vicinity of an Alfv\'en surface, where the advection velocity equals the
Alfv\'en speed. This phenomenon may also have interesting consequences on the
angular momentum redistribution.  

\section*{Acknowledgements}
We thank Henrik Latter and Benjamin Favier for helpful 
discussions and/or comments on the manuscript. We also thank the referee, Thierry Foglizzo, for his insightful report that helped improve the manuscript. JG acknowledges support from the STFC and the Max-Planck-Princeton Center for Plasma Physics. RF acknowledges support from the University of California Office of the President, and from
NSF grants AST-0807444 and AST-1206097.

\appendix

%%%%%%%%%%%%%%%%%%%%%%%%%%%%%%%%%%%%%%%%%%%%%%%%%%%%%%%%%%%%%%%%%%%%%%%%%%%%%%%

\section{The angular momentum flux decomposed into wave contributions}
	\label{sec:decomposition}

The perturbations associated to a SASI mode can be described physically as a
superposition of several kinds of waves: two acoustic waves propagating up and
down, and an advected wave composed of vorticity and entropy perturbations. In
this Appendix we use this decomposition in order to explain the radial profile
of the Reynolds stress associated to a SASI mode, in particular the
oscillations observed in Section~\ref{sec:r2} in the case of higher frequency
harmonics. Strictly speaking, this decomposition requires the use of the WKB
approximation and is only valid for the high frequency harmonics of SASI
\citep{foglizzo07,guilet12}. In order to get a physical understanding in a
cleaner setup, we use a simpler model where the wave decomposition is valid
without any approximation: the planar toy model described by
\citet{foglizzo09}. We only give here a brief description of the aspects
of the model necessary to understand the present analysis, and the reader is
referred to \citet{foglizzo09} for a more complete description of the toy model,
the equations governing it, and the numerical method used to compute the linear
eigenmodes. 

In this model a supersonic flow along the $z$ direction is decelerated through
a shock located at $z=1$. The subsonic flow below the shock is uniform until it
reaches a localised gravity step located around $z=0$ with a width $H_\nabla$.
The vertical axis $z$ is analogous to the radial direction in the core
collapse, while the transverse $x$ and $y$ directions are analogous to the
angular directions $\varphi$ and $\theta$. The $x$ and $y$ directions are here
equivalent and we only consider modes with an $x$ dependence, which is
considered the analog of the azimuthal direction. The linear momentum in the
$x$ direction is then the analog of the angular momentum in spherical geometry.
The perturbation of physical variables take the form:
\begin{equation}
\delta A(x, z, t) = Re\left(\tilde{\delta A(z)}e^{i(k_x x - \omega t)}  \right)
\end{equation}

A surface integration (analogous to the integration over a spherical surface
performed in the rest of the paper) is done on a planar surface at constant $z$ 
over the whole horizontal extent of the box. We define a surface average of the
Reynolds stress (describing the transport in the $z$ direction of momentum in
the $x$ direction) as:
\begin{eqnarray}
T_{Rey} &\equiv& \frac{1}{L_x L_y} \int_0^{L_x}\int_0^{L_y}\rho v_x v_z \, \dd x\dd y \\
T_{Rey} &=& \frac{\rho_0}{2}Re\left(\tilde{\delta v_x}\tilde{\delta v_z^*}\right)
\end{eqnarray}
Similarly to the angular momentum in the spherical case, the surface averaged
linear momentum density in the x direction is then related to the Reynolds
stress by:
\begin{equation}
P_x = -\frac{T_{Rey}}{v_{z0}} + \frac{e^{-2\omega_{i} \tau_{\rm adv}}}{v_{z0}}\int_{\rsh}^r\frac{2\omega_i e^{2\omega_{i} \tau_{\rm adv}}}{v_{z0}}T_{Rey}\, \dd r. 
\end{equation}

The velocity perturbation of a mode can be decomposed into wave contributions
in the following way:
\begin{equation}
\delta v = \delta v^+ + \delta v^- + \delta v^{\rm vort}, 
\end{equation}
where the superscripts $+$ and $-$ refer to the acoustic waves propagating down
and up respectively, and the superscript $\rm vort$ refers to the shear (or
vorticity) wave. Each of the waves has a vertical structure described by a wave
vector $k_z$, for example for acoustic waves:
\begin{equation}
\tilde{\delta A^{\pm}}(z) = \tilde{\delta A_0^\pm}e^{ik_z^\pm z}.
	\label{eq:vert_structure}
\end{equation}

The Reynolds stress can then be written as:
\begin{eqnarray}
T_{\rm Rey} & = &  \frac{\rho_0}{L_x L_y} \int_0^{L_x}\int_0^{L_y}\left\lbrack \delta v_x^+\delta v_z^+ + \delta v_x^-\delta v_z^- + \delta v_x^{\rm vort}\delta v_z^{\rm vort} \right. \nonumber \\
&& \left. + \delta v_x^+\delta v_z^{\rm vort} + \delta v_x^{\rm vort}\delta v_z^+ + \delta v_x^{-}\delta v_z^{\rm vort} + \delta v_x^{\rm vort}\delta v_z^{-}  \right.  \nonumber \\
&& \left. + \delta v_x^+\delta v_z^- + \delta v_x^-\delta v_z^+ \right\rbrack\,\dd^2 s.
	\label{eq:TRey_decomposition}
\end{eqnarray}
The first three terms are the individual contributions of the three waves
(denoted as $T_{Rey}^\pm$ and $T_{Rey}^{\rm vort}$). The last six terms are
cross terms due to the non linear interaction between the waves. Let us define:
\begin{eqnarray}
T_{\rm Rey}^{+-} &\equiv&  \frac{\rho_0}{L_x L_y} \int_0^{L_x}\int_0^{L_y}\left\lbrack \delta v_x^+\delta v_z^- + \delta v_x^-\delta v_z^+\right)\, \dd^2 s    \label{eq:def_TRey_+-} \\
T_{\rm Rey}^{+\rm vort} &\equiv&   \frac{\rho_0}{L_x L_y} \int_0^{L_x}\int_0^{L_y}\left\lbrack \delta v_x^{\rm vort}\delta v_z^+ + \delta v_x^+\delta v_z^{\rm vort}\right)\, \dd^2 s \\
T_{\rm Rey}^{-\rm vort} &\equiv&  \frac{\rho_0}{L_x L_y} \int_0^{L_x}\int_0^{L_y}\left\lbrack \delta v_x^{\rm vort}\delta v_z^- + \delta v_x^-\delta v_z^{\rm vort}\right)\, \dd^2 s
\end{eqnarray}
These cross terms do not generally vanish (although some may in certain cases)
and depend on the relative phase between the different waves. The total
Reynolds stress is then written as the sum of six terms:
\begin{equation}
T_{Rey} = T_{Rey}^{+} +T_{Rey}^{-} + T_{Rey}^{\rm vort} + T_{\rm Rey}^{+-} + T_{\rm Rey}^{-\rm vort} +T_{\rm Rey}^{+\rm vort}
\end{equation}

Next we assess the relative importance of these different contributions, as well
as their sign and vertical profile.
We start by general considerations on each of these
contributions separately (Sections~\ref{sec:acoustic} and \ref{sec:vorticity})
and then describe the wave decomposition below the perturbed shock and the
resulting vertical profile of Reynolds stress between the shock and the
coupling region (Section~\ref{sec:toymodel_vertical_profile}).

\subsection{Acoustic waves}
	\label{sec:acoustic}
Consider a planar acoustic wave with a wave vector ${\bf k} = k_x{\bf u_x} +
k_z{\bf u_z} $ and velocity amplitude $\delta \tilde{v}$. Its velocity
perturbation is given by:
\begin{equation}
\delta {\bf v} = \delta v \frac{\bf k}{k} 
\end{equation}
The Reynolds stress is then (assuming $k_z$ is a real number) :
\begin{equation}
T_{\rm Rey}^{\rm ac} = \frac{\rho_0}{2} |\tilde{\delta v}|^2 \frac{k_x k_z}{k^2} 
	\label{eq:TRey_ac}
\end{equation}
This means that an acoustic wave propagating up ($k_z > 0$) creates an upwards
flux of momentum oriented with the same sign as $k_x$, while a wave propagating
down ($k_z<0$) creates a flux with the opposite sign. The intensity of this
flux is maximum when $k_x=k_z$, i.e. when the wave propagates in an oblique
direction with an angle of $45 ^\circ$ with the vertical. The reason is that large velocities in both horizontal and vertical directions are needed to transport momentum efficiently.

In the presence of advection, the vertical wave number can be expressed as a
function of the frequency $\omega$ and the horizontal wave number as:
\begin{equation}
k_z^{\pm} = \frac{\omega}{c}\frac{\M\mp\mu}{1-\M^2},
\end{equation}
where $\mu$ is defined by $\mu^2 = 1-k_x^2c^2(1-\M^2)/\omega^2$. Assuming that $\omega$ and $\mu$ are real (i.e. the wave is not evanescent), the Reynolds stress associated to an individual wave (+ \emph{or} -) is then:
\begin{eqnarray}
T_{\rm Rey}^{\pm} &=&  \frac{\rho_0}{2} |\tilde{\delta v}|^2 \frac{\omega k_x c (\M\mp\mu)(1- \M^2)}{k_x^2c^2(1-\M^2)^2+\omega^2(\M\mp\mu)^2}. \label{eq:TRey_acbis}
\end{eqnarray}

The cross term coming from the interaction between the two acoustic waves $+$ and $-$ can be written:
\begin{equation}
T_{\rm Rey}^{+-} =   \frac{\rho_0}{2} Re\left(\tilde{\delta v^+}\tilde{\delta v^-}^*\right) \frac{k_x(k_z^+ + k_z^-)}{k^+k^-},
\end{equation}
where $k^\pm = \sqrt{k_x^2 + k_z^{\pm2}}$. In the absence of advection
($\M=0$), then $k_z^+ = -k_z^-$ (for a wave of a given frequency) and $T^{+-}
=0$: the two terms in Equation~(\ref{eq:def_TRey_+-}) cancel each other. In
this special case, the Reynolds stress of the two superposed acoustic waves is
the sum of the individual contributions. In the presence of advection however,
the cross term does not vanish but may be expected to be small for small Mach
numbers: 
\begin{equation}
T_{\rm Rey}^{+-} =    \rho_0 Re\left(\tilde{\delta v^+}\tilde{\delta v^-}^*\right) \frac{k_x}{k^+k^-}\frac{\omega}{c}\frac{\M}{1-\M^2}.
\end{equation}

Note that the term $Re\left(\tilde{\delta v^+}\tilde{\delta v^-}^*\right)$ has the following vertical structure:
\begin{equation}
Re\left(\tilde{\delta v^+}\tilde{\delta v^-}^*\right) = Re\left(\tilde{\delta v^+_0}\tilde{\delta v^-_0}^* e^{i(k_z^+-k_z^-)z)}\right),
\end{equation}
i.e. it oscillates with a wave number $k_z^\pm = k_z^+ - k_z^-$. Contrary to
the Reynolds stress contribution of an individual wave which does not depend on
its phase and therefore does not oscillate, the cross term depends on the
relative phase of the two waves and therefore displays vertical oscillations
with a wave vector that is the difference between the wave vectors of the two
waves.

\subsection{Shear wave}
	\label{sec:vorticity}
A shear wave with a wave vector ${\bf k} = k_x{\bf u_x} + k_z{\bf u_z}$ (with
$k_z = \omega/v_0 < 0$) and velocity amplitude $\delta \tilde{v} $ has the
following velocity vector:
\begin{equation}
{\bf \delta v} = \delta \tilde{v} \, {\bf u_y}\times \frac{\bf k}{k}.
\end{equation}
The Reynolds stress of this wave is then:
\begin{eqnarray}
T_{\rm Rey}^{\rm vort} &=& - \frac{\rho_0}{2} |\delta \tilde{v}|^2 \frac{k_x k_z}{k^2}  \label{eq:TRey_vort1} \\
&=&  - \frac{\rho_0}{2} |\delta \tilde{v}|^2 \frac{1}{\frac{\omega}{k_x v} + \frac{k_xv}{\omega}} \label{eq:TRey_vort2}.
\end{eqnarray}
Note that this corresponds to a momentum flux opposite to that of an acoustic
wave with the same wave vector and velocity amplitude (compare
Equations~(\ref{eq:TRey_ac}) and (\ref{eq:TRey_vort1})). The reason is that the
velocity associated with an acoustic wave is parallel to its wave vector, while
the velocity of a shear wave is perpendicular to its wave vector (because ${\bf
\nabla . v} = i{\bf k.v} =0 $). As $k_z$ is negative, the Reynolds stress
carried by the shear wave has the same sign as $k_x$. For a given velocity
amplitude, the optimum orientation of the wave vector is again $45^\circ$ with
respect to the vertical. 

The cross term between the vorticity wave and the acoustic waves can be written as:
\begin{equation}
T_{\rm Rey}^{\pm vort} = \frac{\rho_0}{2}Re\left( \delta \tilde{v}^{\rm vort}\delta\tilde{v}^{\pm*}\right) \frac{k_z^{\vort}k_z^\pm - k_x^2}{k^{\pm}k^{\vort}},
	\label{eq:TRey_vortac}
\end{equation}
which oscillates in the vertical direction with a wave number $k_z^{\vort\pm} = k_z^{\vort}-k_z^\pm$.

\subsection{Vertical profile of the Reynolds stress}
	\label{sec:toymodel_vertical_profile}

\begin{figure}
\centering
 \includegraphics[width=\columnwidth]{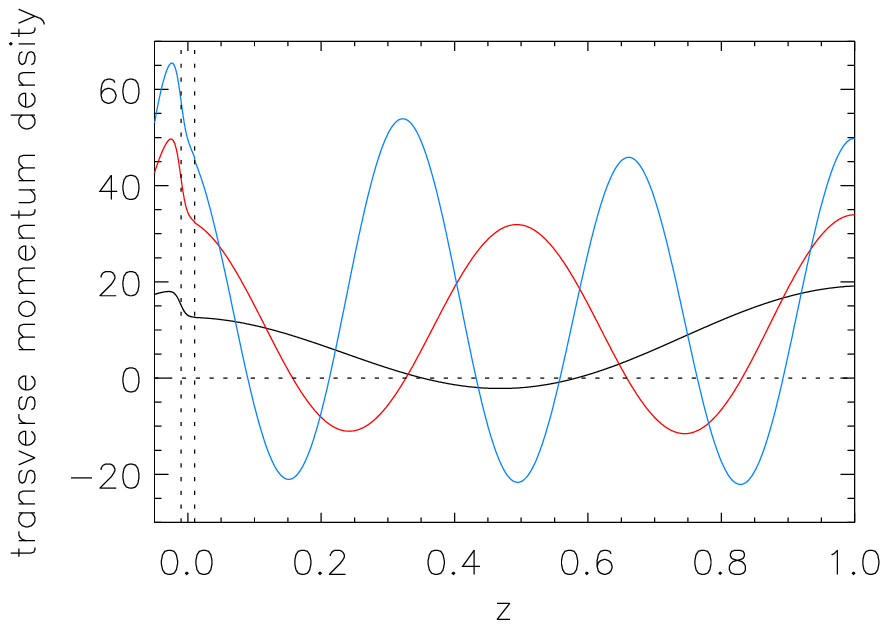}
 \caption{Vertical profile of the transverse momentum in the planar toy model
of \citet{foglizzo09} with the parameters $L_x = 4$, $n_x=1$, $c_{\rm
in}^2/c_{\rm out}^2 = 0.75$, $H_\nabla = 0.02 H$, and a strong shock ($\M_1
\rightarrow \infty$). The fundamental mode is shown with a black line, the
first higher frequency harmonic with a red line, and the second harmonic with
a blue line. }
             \label{fig:toymodel_Reynolds}%
\end{figure}

\begin{figure*}
\centering
 \includegraphics[width=0.66\columnwidth]{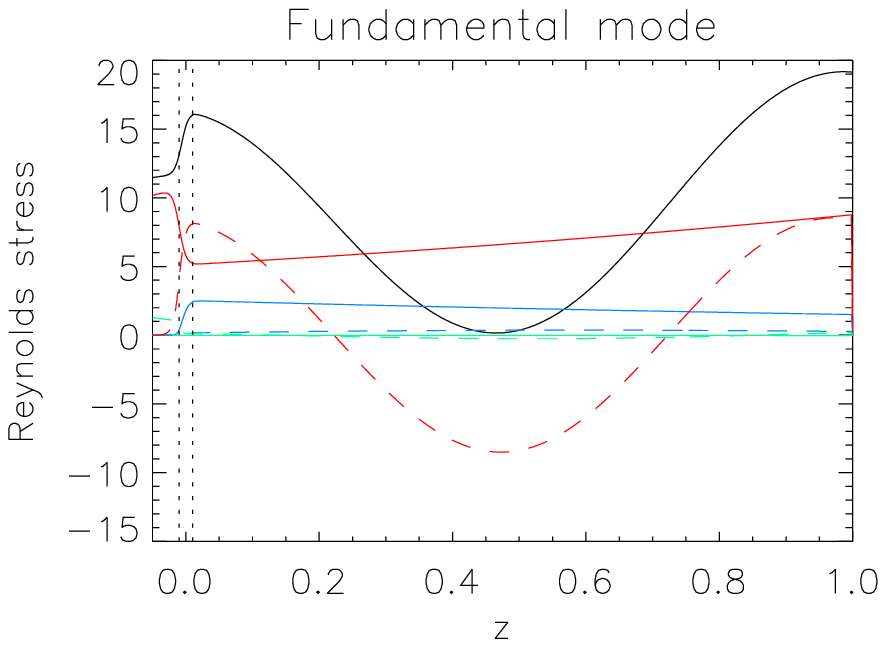}
  \includegraphics[width=0.66\columnwidth]{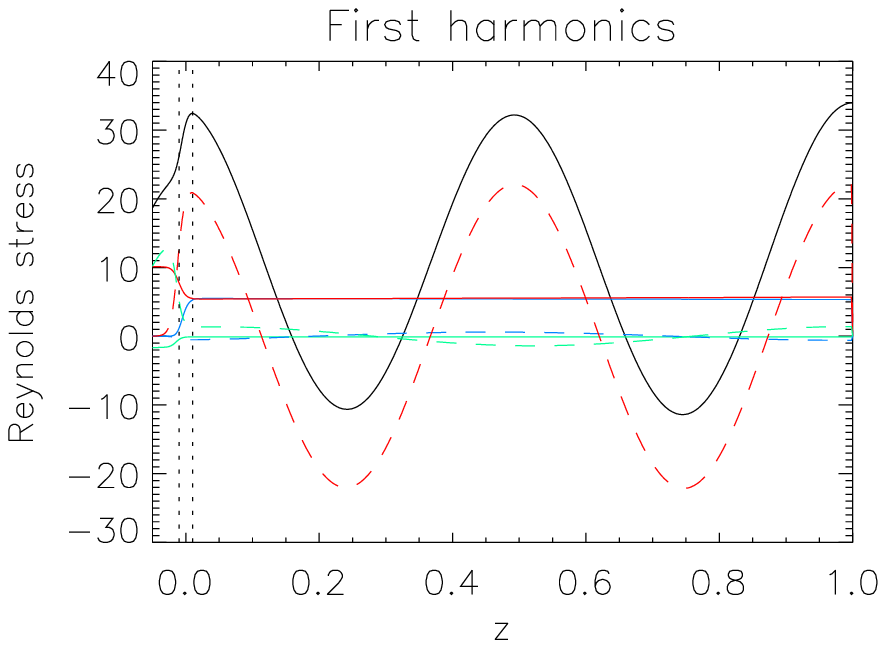}
  \includegraphics[width=0.66\columnwidth]{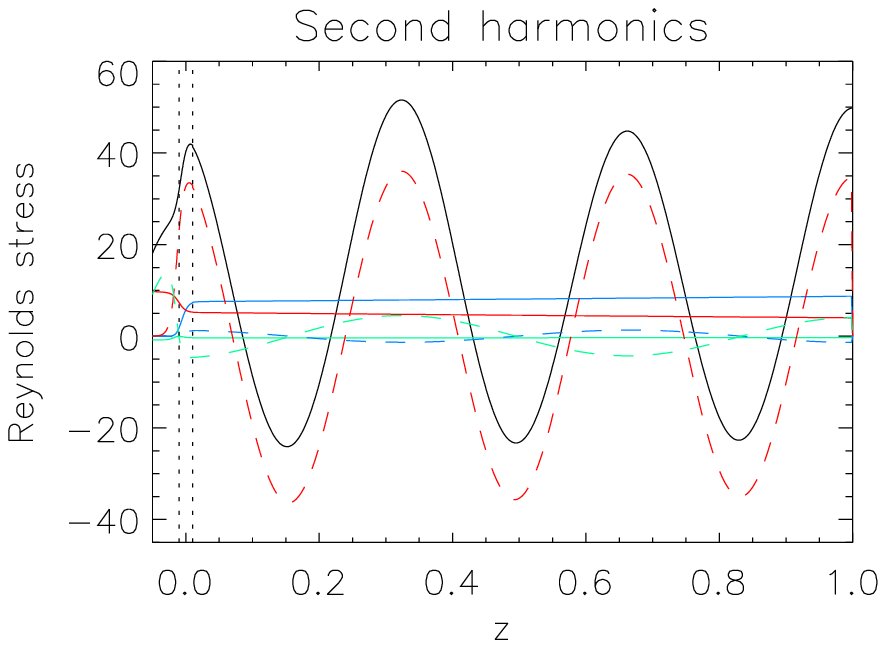}
 \caption{Decomposition of the Reynolds stress into wave contributions in the
planar toy model of \citet{foglizzo09}. The parameters are the same as those
used in Figure~\ref{fig:toymodel_Reynolds}. The different lines show the total
Reynolds stress $T_{\rm Rey}$ (black line), the contributions of individual
waves : the vorticity wave $T_{\rm Rey}^{\rm vort}$ (red full line), the
acoustic wave propagating up $T_{\rm Rey}^-$ (blue full line), and the acoustic
wave propagating down $T_{\rm Rey}^+$ (green full line), and finally the
contributions due to the interaction between two types of waves : the vorticity
wave and the acoustic wave propagating up $T_{\rm Rey}^{-\rm vort}$ (red dashed
line), the vorticity wave and the acoustic wave propagating down $T_{\rm
Rey}^{+ \rm vort}$ (green dashed line), and the two acoustic waves $T_{\rm
Rey}^{+-}$ (blue dashed line).  The three panels show three different modes
with increasing frequencies : fundamental mode (left panel), first higher
frequency harmonics (middle panel), and second higher frequency harmonics
(right panel). The vertical dotted lines show the extent of the potential
jump.}
             \label{fig:toymodel_decomposition}%
\end{figure*}

The vertical profiles of transverse momentum induced by three modes with a
transverse structure $n_x=1$ are shown in Figure~\ref{fig:toymodel_Reynolds}:
the fundamental mode (black line), the first higher frequency harmonics (red
line) and the second higher frequency harmonics (blue line). They show
oscillations that are very similar to the radial oscillations observed in the
spherical model in Section~\ref{sec:r2}. The fundamental mode makes one
oscillation, the first harmonics two oscillations and the third harmonics three
oscillations between the shock and the potential jump. To understand this
feature, we show in Figure~\ref{fig:toymodel_decomposition} the decomposition
of the Reynolds stress into the six contributions coming from the three types
of waves and the interaction between them. There are three significant
contributions: the vorticity wave (red full line), the acoustic wave
propagating up (blue full line) and the interaction between these two waves
(red dashed line). The other three contributions are significantly smaller:
the acoustic wave propagating down and its interaction with the two other
waves. This is consistent with the fact that SASI is caused by an
advective-acoustic cycle in which the advected vorticity wave and the acoustic
wave propagating up play a dominant role, while the acoustic wave propagating
down plays only a minor role. The individual contributions of the vorticity and
acoustic waves do not oscillate and are both positive (for $k_x>0$) as shown in
the last two subsections. The oscillations in the Reynolds stress profile are
due to the interaction between the vorticity wave and the acoustic wave
propagating up. As shown in the previous section, the Reynolds stress resulting
from this interaction $T_{\rm Rey}^{-\rm vort}$ oscillates with a wave vector :
$\omega/v - k_z^-$. If there is no phase shift at the couplings (at the shock
and in the gradient), then the phase relation determining the frequency of
avective-acoustic modes is : $(\omega/v - k_z^-)H = 2n_z\pi$ (this is
equivalent to Equation~34 of \citet{guilet12}), where $H$ is the distance
between the shock and the potential jump and $n_z$ is an integer number
defining the mode considered ($n_z=1$ for the fundamental mode, $n_z=2$ for the
first harmonics, $n_z=3$ for the second harmonics). This therefore explains the
number of oscillations observed in the Reynolds stress profile.    

The relative importance of the different  contributions can be determined
analytically by using the boundary conditions at the shock. For simplicity we
restrict our analysis to the case of a strong shock and an eigenfrequency
$\omega$ which is real and such that acoustic waves are not evanescent (i.e.
$\mu^2>0$). Neglecting the growth rate leads to an error which is less than
$1\%$ of the total Reynolds stress. The velocity perturbations below the shock
and the resulting Reynolds stress can be written as :
\begin{eqnarray}
\delta v_{x\sh} &=&   \frac{2}{\gamma -1}ik_xv_\sh \Delta z  \\ 
\delta v_{z\sh} &=&  -  \frac{2}{\gamma +1}i\omega \Delta z \\ 
T_{Rey} &=& -\frac{2}{\gamma^2 -1}\rho_0 v_\sh k_x\omega\Delta z^2 
\end{eqnarray}

The $y$-component of the vorticity created by the shock oscillations is \citep{foglizzo09}:
\begin{equation}
\delta w_\sh = - \frac{4}{\gamma^2-1} \omega k_x \Delta z,
\end{equation}
and the contribution of the shear wave to the Reynolds stress at the shock is
then:
\begin{equation}
T_{\rm Rey}^{\rm vort} = \frac{4}{\gamma^2-1} \frac{1}{(\frac{\omega}{k_x v} + \frac{k_xv}{\omega})^2} T_{Rey}.
\end{equation}
The fraction of the total Reynolds stress contributed by the vorticity wave is
maximum if $\frac{\omega}{k_xv} = 1$, i.e. if the vorticity is inclined by an
angle of $45^\circ$ with respect to the vertical (recall that $k_z =
\omega/v$). When $\frac{\omega}{k_xv} $ goes to zero or infinity this fraction
goes to zero. For the fundamental mode and the parameters used in
Figures~\ref{fig:toymodel_Reynolds} and \ref{fig:toymodel_decomposition},
$\omega/(k_xv) \simeq 3$ and $T^{\rm vort} \simeq 0.5 T_{\rm Rey}$. In the case
of higher frequency harmonics the vorticity wave contributes less to the total
Reynolds stress in agreement with Figure~\ref{fig:toymodel_decomposition}. This
comes from the fact that, when the frequency increases, the vertical component
of the wave vector of the shear wave becomes more and more dominant. Since the
velocity perturbation associated to a shear wave is perpendicular to its wave
vector, the velocity is then mostly horizontal. As a result the vertical
component of the velocity associated to the shear wave decreases, thus
decreasing the associated Reynolds stress. 

In order to extrapolate this result to the spherical model, one can estimate
the equivalent of the parameter $\omega/(k_xv)$ to be $\omega
r_\sh/(\sqrt{l(l+1)} v_\sh)$. For the fundamental mode when $r_*=0.5r_\sh$,
this parameter is approximately $3.5$, therefore rather close to its value for
the fundamental mode of the planar toymodel. We therefore expect that the
vorticity wave contributes to about half of the total Reynolds stress. This
feature can be expected to hold generally for the most unstable mode. Indeed
the most unstable mode corresponds to acoustic waves being close to horizontal
propagation (see Figure~8 and 10 of \citet{guilet12}), i.e. $\omega \simeq k
c_s$ where $k$ is the transverse wave number and $c_s$ is the sound speed.
Therefore the most unstable mode should satisfy $\omega/(k v) \simeq 1/M_\sh
\simeq 3$. 

The pressure perturbation below the shock can be expressed as: 
\begin{equation}
\left(\frac{\delta P}{\gamma P}\right)_\sh = \frac{2}{\gamma}\frac{\gamma -1}{\gamma + 1}\frac{i\omega\Delta z}{v_\sh}.
\end{equation}
This can be decomposed into the two acoustic waves propagating up and down, which have the following pressure and velocity perturbations:
\begin{eqnarray}
\left(\frac{\delta P^\pm}{\gamma P}\right)_\sh &=& \frac{\gamma -1}{\gamma(\gamma + 1)}\frac{i\omega\Delta z}{v_\sh}\left(1 \mp \frac{\mu}{2\M} \right), \\
\delta v^\pm_\sh &=& - \frac{1}{\gamma + 1}\sqrt{\frac{2(\gamma -1)}{\gamma}} i\omega\Delta z\left(1 \mp \frac{\mu}{2\M} \right).
\end{eqnarray}
The Reynolds stress associated to each of these two acoustic waves can then be
obtained by substituting this velocity perturbation into
Equation~(\ref{eq:TRey_acbis}). Note that the velocity amplitude of the acoustic
wave propagating up (noted -) is always larger than that of the acoustic wave
propagating down (noted +): the acoustic wave is damped at the reflection at
the shock as was already shown by \citet{foglizzo09}. This explains why the
three terms involving the acoustic wave propagating down ( $T_{\rm Rey}^{+}$,
$T_{\rm Rey}^{+ \rm vort}$, and $T_{\rm Rey}^{+-}$) play only a very minor role
in the Reynolds stress.

The Reynolds stress from the interaction between the vorticity wave and the
acoustic waves can be found using Equation~(\ref{eq:TRey_vortac}) and the
velocity perturbations of the waves:
\begin{equation}
T_{\rm Rey}^{\pm \rm vort} = \frac{1}{\gamma + 1}\sqrt{\frac{2(\gamma -1)}{\gamma}} \frac{\omega}{v}\frac{\left( 1\mp \frac{\mu}{2\M}\right)\left( \frac{\omega^2}{vc}\frac{\M\mp\mu}{1-\M^2} - k_x^2 \right)}{\left(\frac{\omega^2}{v^2} + k_x^2 \right)\sqrt{\frac{\omega^2}{c^2}\left(\frac{\M\mp\mu}{1-\M^2}\right)^2 + k_x^2}} T_{\rm Rey}
\end{equation}
The contribution from the interaction between the vorticity wave and the
acoustic wave propagating up has the same sign as the total Reynolds stress and
contributes a significant fraction of it: about half for the fundamental mode,
and more than two thirds for the second harmonics. The fact that it contributes
a large fraction of the total stress at high frequency can be understood
qualitatively as follows. At high frequency, the wave vectors of both the
vorticity wave and the acoustic wave are mostly vertical. As a result, the
velocity perturbation of the vorticity wave is mostly horizontal while the
velocity perturbation of the acoustic wave is mostly vertical. Neither of the
waves independently transports efficiently angular momentum because this needs
both horizontal and vertical velocities. But the interaction of the two waves
is very efficient at transporting angular momentum by combining the horizontal
velocity of the vorticity wave and the vertical velocity of the acoustic wave.

\section{Measure of the spiral modes amplitude in the linear phase}
	\label{sec:spiral_mode_amplitude}
In this appendix we describe the method we used in Section~\ref{sec:linear} in
order to compute the amplitude of the two spiral modes $\pm m$ in the linear
phase of SASI. We describe successively two different methods based on a fit of the time-evolution either of the displacement amplitude of the shock (Section~\ref{sec:fit_shock_displacement}) or of the transverse velocities in the postshock flow (Section~\ref{sec:fit_transverse_velocities}). The results of these two methods are then compared in Section~\ref{sec:fit_comparison}.

\subsection{Method using the shock displacement}
	\label{sec:fit_shock_displacement}
\begin{figure}
\centering
 \includegraphics[width=\columnwidth]{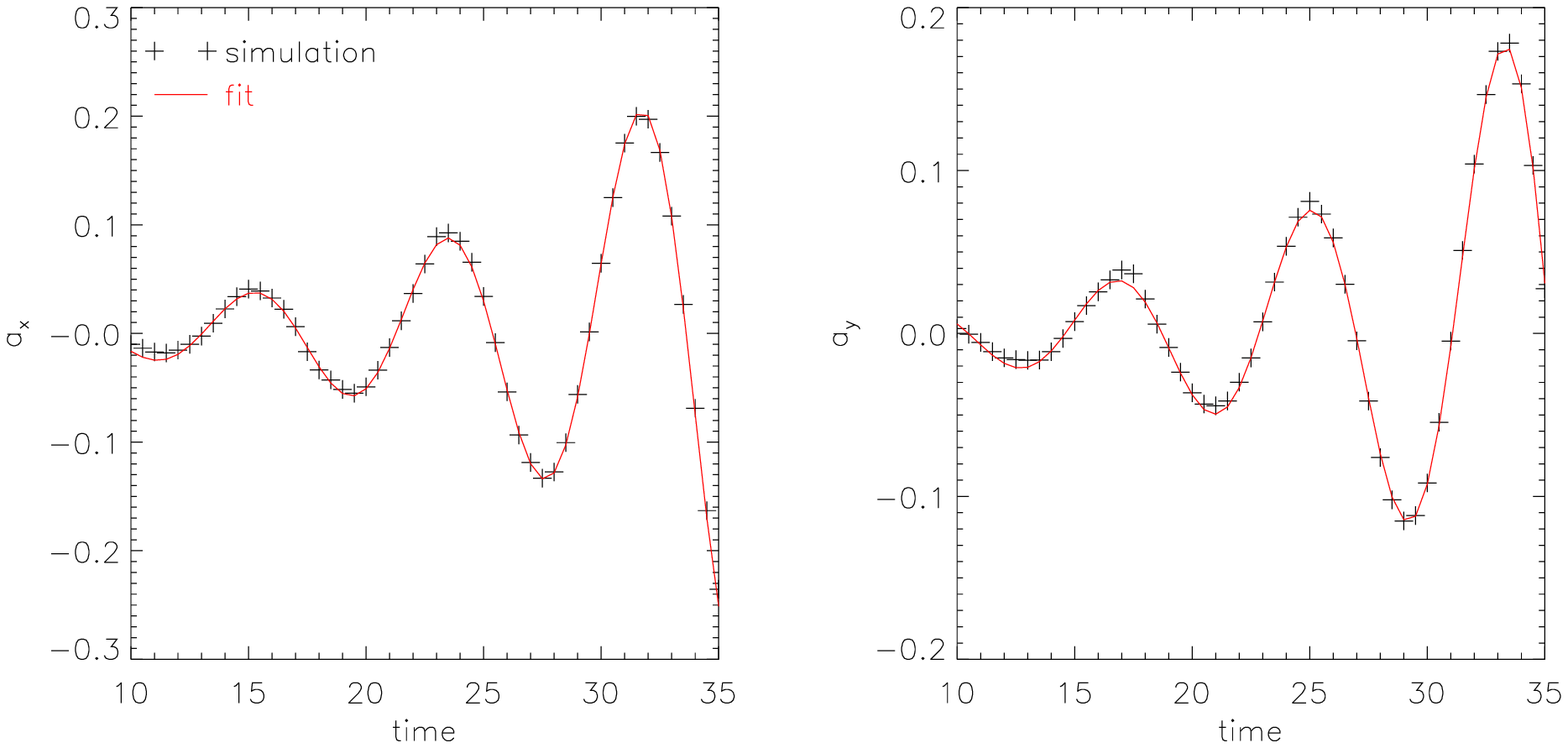}
 \caption{Fit of the amplitude of the shock deformation projected onto
spherical harmonics in the linear phase of SASI. Left panel: along the x axis.
Right panel: along the y axis. The simulation results are shown with black $+$
signs, while the fit is shown with red line.} \label{fig:fit}%
\end{figure}

To be specific, we here focus on the case of spiral modes with
$m=\pm 1$ (Section~\ref{sec:r5} and ~\ref{sec:r2}) but the same method can be
applied to the case of $m=\pm2$ as well (Section~\ref{sec:r6}). The amplitude
of the shock deformation has been projected onto real spherical harmonics along
the x and y axis (we assume that the component on the z axis is negligible)
defined as:
\begin{eqnarray}
Y_{1x} &=& \sqrt{\frac{3}{4\pi}}\sin\theta\cos\phi \\
Y_{1y} &=& \sqrt{\frac{3}{4\pi}}\sin\theta\sin\phi
	\label{eq:real_spherical_harmonics} 	
\end{eqnarray}
(Note that $Y_{1x}$ was noted $Y_1^1$ in \citet{fernandez10}, while $Y_{1y}$
was noted $Y_1^{-1}$.) The time evolution of these spherical harmonics
amplitudes is then fitted using a function of the form:
\begin{equation}
f(t) = A\cos(\omega_r t + \Phi)\exp{\omega_i t},
	\label{eq:fit}
\end{equation} 
which has four parameters: the amplitude $A$, the phase $\Phi$, the frequency
$\omega_r$ and the growth rate $\omega_i$. These are measured for both axis x
and y (Figure~\ref{fig:fit}). As expected, the frequency and growth rate are
the same on both axis within small numerical errors, and we later use the
average of the two values.

These two sloshing modes can be equivalently described as two counter-rotating
spiral modes, which are described using complex spherical harmonics defined as:
\begin{eqnarray}
Y_1^1 &=& -\sqrt{\frac{3}{8\pi}}\sin\theta e^{i\phi} \\
Y_1^{-1} &=& \sqrt{\frac{3}{8\pi}}\sin\theta e^{-i\phi}
	\label{eq:complex_spherical_harmonics} 	
\end{eqnarray}
 
The sloshing mode along the $x$ axis ($\phi=0$) with amplitude $A_x$ and phase
$\Phi_x$ can decomposed into two spiral modes of equal amplitude but different
phase as:
\begin{equation}
A_x\cos(\omega_r t + \Phi_x) e^{\omega_i t} Y_{1x} = Re\left\lbrack\frac{A_x}{\sqrt{2}}e^{-i\phi_x}e^{-i\omega t} \left( -Y_1^{1} + Y_1^{-1} \right)   \right\rbrack
	\label{eq:sloshing_x_decomposition}
\end{equation}

Similarly the sloshing mode along the $y$-axis can be decomposed into two spiral modes of equal amplitude and phase:
\begin{eqnarray}
A_y\cos(\omega_r t + \Phi_y) e^{\omega_i t} Y_{1y} &=& Re\big\lbrack\frac{A_y}{\sqrt{2}}e^{-i(\phi_y-\pi/2)}e^{-i\omega t}  \nonumber \\
&& (Y_1^{1} + Y_1^{-1} )   \big\rbrack
	\label{eq:sloshing_y_decomposition}
\end{eqnarray}

Combining these two expressions, we finally obtain the amplitude and phase of the two spiral modes $m=\pm 1$ as:
\begin{eqnarray}
A_1 e^{-i\phi_1} &=& -\frac{A_x}{\sqrt{2}}e^{-i\phi_x} + \frac{A_y}{\sqrt{2}}e^{-i(\phi_y-\pi/2)}  \\
A_{-1} e^{-i\phi_{-1} } &=& \frac{A_x}{\sqrt{2}}e^{-i\phi_x} + \frac{A_y}{\sqrt{2}}e^{-i(\phi_y-\pi/2)} 
	\label{eq:spiral_amplitude_m1}
\end{eqnarray}
Note that here the amplitude of the sloshing mode is expressed in terms of real
spherical harmonics, while that of the spiral modes is expressed in terms of
complex spherical harmonics which have a different normalisation (by a factor
$1/\sqrt{2}$).

\subsection{Method using transverse velocities}
	\label{sec:fit_transverse_velocities}
To check our results, we have computed the amplitude of spiral modes with an alternative
method involving the transverse velocities of the flow. To this end we
make use of the quantity $\delta A$ defined in \citet{foglizzo07}:
\begin{eqnarray}
\label{eq:dA_definition}
\delta A & \equiv & \frac{r}{\sin\theta}\left[\frac{\partial }{\partial\theta}\left(\sin\theta \delta v_\theta\right)
					  + \frac{\partial}{\partial \phi}\delta v_\phi \right]\\
&=&\frac{1}{i\omega}\left[\delta K - \ell(\ell+1)f\right],
\end{eqnarray}
which vanishes for a spherically symmetric flow, and which has been previously
used to compute the perturbation amplitude by \citet{scheck08}.

The transverse velocities have the dependence of a vector spherical harmonic,
\begin{equation}
\delta \mathbf{v}_\perp = \delta v_\theta\hat\theta + \delta v_\phi\hat\phi =  \delta \tilde{v}_\perp e^{-i\omega t}\mathbf{\Psi}_{\ell m}(\theta,\phi),
\end{equation}
with
\begin{equation}
\mathbf{\Psi}_{\ell m} = r\nabla Y_\ell^m = \hat\theta\,\frac{\partial}{\partial\theta}Y_\ell^m 
                + \hat\phi\,\frac{1}{\sin\theta}\frac{\partial}{\partial \phi}Y_\ell^m,
\end{equation}
where $\hat \theta$ and $\hat \phi$ are the unit coordinate vectors in the polar and
azimuthal directions, respectively. Equation~(\ref{eq:dA_definition}) thus implies that
$\delta A$ is proportional to a scalar spherical harmonic $Y_\ell^m$, with amplitude
\begin{equation}
\label{eq:dA_amplitude}
\delta \tilde{A} = -\ell(\ell+1) r \delta \tilde{v}_{\perp},
\end{equation}
and also proportional to the shock displacement amplitude $\Delta \tilde{r}$. 

To find $\Delta \tilde{r}$ from the simulation, we first project the transverse
velocity field into the appropriate vector spherical harmonic to obtain a 
coefficient
\begin{equation}
\tilde{v}_{\perp, \textrm{ sim}}(r,t) = \frac{1}{\ell(\ell+1)}\int \mathbf{v}_\perp(r,\theta,\phi,t) 
				\cdot \mathbf{\Psi}^*_{\ell m}\,\dd \Omega.
\end{equation}
$\Psi_{\ell m}$ is computed from real spherical harmonics along $x$ and $y$ axis, and the velocity coefficient is therefore real. The time evolution of these velocity amplitudes at a chosen radius is then fitted using the same method as in last subsection. The complex amplitude of spiral modes thus obtained $\Delta \tilde{v}$ is then converted into a complex shock displacement amplitude by using equation~(\ref{eq:dA_amplitude}) in the following way
\begin{equation}
\label{eq:amplitude_alt_equation}
\Delta \tilde{r} = \frac{-\ell(\ell+1)\,r\, \Delta \tilde{v}}
     {\delta \tilde A},
\end{equation}
where the complex amplitude $\delta \tilde{A}$ is obtained from the linear stability analysis with a unit shock displacement at phase zero.

\subsection{Comparison of the two methods}
	\label{sec:fit_comparison}
We have applied the two above methods to the linear growth phase of model R5\_L11\_HR (analysed in Section~\ref{sec:r5}). The eigenfrequencies obtained from the fit match the linear analysis to within $0.5\%$ for the oscillation frequency and to within $5\%$ for the growth rate. The first method gives the following amplitudes of the two spiral modes $m=1$ and $m=-1$ at time $t=30$: $A_1=0.206$ and $A_{-1}=0.051$. This can be compared to the amplitude measured using transverse velocities at radius $r_1 = 0.6$: $A_1 = 0.232$ and $A_{-1} = 0.052$, and at radius $r_2 = 0.8$: $A_1 = 0.229$ and $A_{-1}=0.055$. The two measures using the second method at different radii are in good agreement with each other, while they are $\sim10\%$ larger than the result of the first method. In Section~\ref{sec:linear}, we have chosen to use the result of the second method because it provides a better match to the radial profiles of linear perturbations. 

For model R6\_L22\_P2 (analysed in Section~\ref{sec:r6}), the first method gives: $A_2=0.180$ and $A_{-2} = 0.022$. The second method using $r_1=0.7$ gives: $A_2=0.177$ and $A_{-2} = 0.017$, while using $r_2=0.8$ we obtain:  $A_2=0.185$ and $A_{-2} = 0.021$. The agreement between the two methods is here better than for model R5\_L11\_HR in spite of the lower resolution.

\bibliography{supernovae}

\bsp
\label{lastpage}

\end{document}